\documentclass[aps,prl,twocolumn,footinbib,amsmath,amssymb,amsfonts,superscriptaddress]{revtex4-2}

\usepackage{color,framed,graphicx}
\usepackage{tikz}
\usetikzlibrary{arrows,decorations.pathmorphing,decorations.markings}
\usepackage{amsmath, amssymb, amsfonts,mathrsfs, amsthm}
\usepackage{physics}
\usepackage{lipsum}
\usepackage[hypertexnames=false]{hyperref}
\usepackage{enumerate}
\usepackage{array}
\usepackage{verbatim}

\usepackage{multibib}

\begin{document}

\title{Universal Spin Screening Clouds in Local Moment Phases}

\author{Minsoo L. Kim}
\affiliation{Department of Physics, Korea Advanced Institute of Science and Technology, Daejeon 34141, Korea}

\author{Jeongmin Shim}
\affiliation{Department of Physics, Korea Advanced Institute of Science and Technology, Daejeon 34141, Korea}

\author{H.-S. Sim}\email[]{hssim@kaist.ac.kr}
\affiliation{Department of Physics, Korea Advanced Institute of Science and Technology, Daejeon 34141, Korea}

\author{Donghoon Kim}\email[(Current affiliation: RIKEN) ]{donghoon.kim@riken.jp}
\affiliation{Department of Physics, Korea Advanced Institute of Science and Technology, Daejeon 34141, Korea}

\date{\today}

\begin{abstract}

When a local impurity spin interacts with conduction electrons whose density of states (DOS) has a (pseudo)gap or diverges at the Fermi energy, a local moment (LM) phase can be favored over a Kondo phase.
Theoretically studying quantum entanglement between the impurity and conduction electrons, we demonstrate that conduction electrons form an ``LM spin cloud'' in general LM phases, which corresponds to, but has fundamental difference from, the Kondo cloud screening the impurity spin in the Kondo phase.
The LM cloud algebraically decays over the distance from the impurity when the DOS has a pseudogap or divergence, and exponentially when it has a hard gap.
We find an ``LM cloud length'', a single length scale characterizing a universal form of the LM cloud.
The findings are supported by both of analytic theories and numerical computations.

\end{abstract}
\maketitle

The Kondo effect arises when conduction electrons screen a local impurity spin-1/2~\cite{Hewson97}, leading to the formation of a spin singlet state~\cite{Yoshida66,Kim21,Shim23,Yoo18,Lee15} having the maximal entanglement between the impurity spin and the conduction electrons.
The electrons form a spatial structure around the impurity, the Kondo screening cloud \cite{Affleck09}. 
The cloud has a universal spatial distribution scaled by a single length $\xi_{\text{K}}$, the Kondo cloud length.
It was experimentally observed~\cite{Borzenets20}.
A condition for the Kondo effect is that the conduction electrons have a finite density of states (DOS) at the Fermi level.

When the DOS exhibits a (pseudo)gap or divergence at the Fermi level, a local moment (LM) phase~\cite{Withoff90, Mitchell13, Chen98} competes with the Kondo effect.
It is favored when the interaction between the impurity and conduction electrons is sufficiently weak.
The LM phases appear in diverse materials, including superconductors~\cite{Satori92, Moca21}, semimetals~\cite{Fritz13}, and heavy-fermion compounds~\cite{Coleman15, Si01}. 

It has been believed that the impurity-spin screening by conduction electrons vanishes in the LM phase~\cite{Withoff90,Vojta06}.
The background argument is that the LM phase corresponds to the fixed point describing the decoupling between the impurity and conduction electrons~\cite{Withoff90,Vojta06}.
Indeed, thermodynamic quantities such as impurity entropy and spin susceptibility behave as if no screening happens at zero temperature \cite{Vojta06, Wagner18}.
This fixed point corresponds to the high-temperature phase of the Kondo effects.
The vanishing impurity screening~\cite{Kim21,Shim23} at the high temperature seems to support the belief.

However, this belief is challenged by a recent theory by Moca et al.~\cite{Moca21}.
It shows
the formation of a spin cloud screening an impurity spin in the Yu-Shiba-Rusinov phase of a superconductor.
The phase is an LM phase, as the superconductor is gapped.
The result suggests to study whether a spin cloud appears in general LM phases, 
and how a spin cloud (if exists) exhibits universal features and differs from the Kondo cloud.
  
In this work, we show that an ``LM spin screening cloud'' appears in general LM phases, studying all possible forms (pseudogap, hard gap, diverging) of the DOS.
We consider the entanglement~\cite{Shim23,Kim21} between the impurity spin and conduction electrons,
and derive a relation between the entanglement and the expectation value of the impurity spin.
The spatial distribution of the LM cloud is characterized by an emergent length, which we call ``LM cloud length''.
Its spatial decay over the distance from the impurity follows a universal function of the ratio between the distance and the LM cloud length.

Unlike the Kondo cloud, the LM cloud screens the impurity spin only partially, as its entanglement with the impurity is not maximal, and the universal function describing its spatial decay depends on the DOS. When the DOS exhibits a pseudogap or diverges, the universal function follows a power law. We derive the power law by developing fixed point analysis~\cite{Wilson75, Buxton98}. It agrees with numerical renormalization group (NRG) results~\cite{Wilson75, Shim18}.
When the DOS has a hard gap, on the other hand, the universal function follows an exponential decay.
For a sufficiently large interaction strength, the decay length, namely the LM cloud length, is determined by the inverse of the gap, as in a superconductor~\cite{Moca21}.
Otherwise, the length nontrivially depends on the gap and the DOS form.
Density matrix renormalization group (DMRG) calculations~\cite{Schollwock11} support this.

{\it Kondo model with various DOSs.---}
We study the Kondo model, $H_{\text{K}} =\sum_{\mathbf{k},\sigma} {\epsilon}_{\mathbf{k}}\,{c_{\mathbf{k}\sigma}^\dagger}{c_{\mathbf{k}\sigma}}+J\vec{S}_{\text{imp}}\cdot\vec{S}_{\text{bath}}$. $c^\dagger_{\mathbf{k}\sigma}$ creates an electron of energy $\epsilon_{\mathbf{k}}$ and spin $\sigma$ at
momentum $\mathbf{k}$. $\vec{S}_{\text{imp}}$ is the impurity spin-1/2 operator. $\vec{S}_{\text{bath}}$ is the electron spin at the impurity position. $J$ describes Kondo coupling.
We consider the energy-dependent DOS $\rho(\epsilon)$, satisfying $\rho(\epsilon) = \rho(-\epsilon)$. 
As the LM phase emerges only when the DOS at Fermi level either vanishes or diverges, we classify the DOS into the three cases.
The first is a power-law DOS involving a pseudogap at Fermi level,
\begin{equation}
	\label{eq1}
	\rho(\epsilon) = \frac{1+r}{2D} (|\epsilon/D|)^r \Theta(D - |\epsilon|)
\end{equation}
with $r>0$. $D$ is the bandwidth.
The second is a diverging DOS following the same power-law of Eq.~\eqref{eq1} but with $r < 0$.
The last hard-gap DOS vanishes in a gap $\Delta$, 
\begin{equation}
	\label{eq2}
	\rho(\epsilon) = f(\epsilon) \Theta(|\epsilon| - \Delta) \Theta(D - |\epsilon|),
\end{equation}
with an arbitrary even function $f(\epsilon)$. To model an electron bath of a hard gap, we employ the Su-Schrieffer-Heeger (SSH) model \cite{Su79,Supp}.

The ground-state phase depends on the DOS and the coupling $J$. For the pseudogap DOS with $0<r<0.5$, the system is in the LM phase at $J < J_{\text{c}}$, while it is in the Kondo phase at $J > J_{\text{c}}$~\cite{Withoff90, Buxton98}. $J_\text{c}$ is the critical coupling strength. 
When $r \ge 0.5$, it is always in the LM phase.
For the diverging DOS, it depends on whether there is a potential scattering of $H_V = V \sum_{\mathbf{k}, \mathbf{k}', \sigma} c_{\mathbf{k}\sigma}^\dagger c_{\mathbf{k}'\sigma}$.
In the absence of the potential scattering, it is always in the Kondo phase.
In the presence, it is in the asymmetric local moment (ALM) phase~\cite{Mitchell13} at $J<J'_{\text{c}}$,
and in the Kondo phase at $J>J'_{\text{c}}$, with the critical coupling strength $J'_\text{c}$. 
For the hard gap DOS, it is always in the LM phase (in the absence of the potential scattering)~\cite{Chen98, Zalom23}.

{\it Entanglement in LM phases.---}
We consider the entanglement negativity~\cite{Vidal02} $\mathcal{N}=\Vert\hat{\rho}^{\text{T}_{\text{imp}}}\Vert_{1}-1$ between the impurity and conduction electrons, 
as it successfully quantifies the impurity spin screening in Kondo effects~\cite{Yoshida66,Lee15}.
$\hat{\rho}$ is the density matrix of the total system, $\hat{\rho}^{\text{T}_{\text{imp}}}$ is the partial transpose of $\hat{\rho}$ on the impurity spin, and $\Vert\hat{\rho}^{\text{T}_{\text{imp}}}\Vert_{1}$ is the trace norm of $\hat{\rho}^{\text{T}_{\text{imp}}}$.

Non-zero negativity implies non-vanishing spin screening~\cite{Shim18,Kim21,Shim23}.
To see this, we consider a zero-temperature quantity~\cite{defofS} $\mathcal{S} \equiv |\langle\Psi_{\text{GS}}|\vec{S}_{\mathrm{imp}}|\Psi_{\text{GS}}\rangle|$ of the impurity. $|\Psi_{\mathrm{GS}}\rangle$ is any arbitrary pure state in the ground state manifold.
$\mathcal{S}$ is independent of the choice of $|\Psi_{\mathrm{GS}}\rangle$ due to the SU(2) symmetry.
We find~\cite{Supp} the connection between $\mathcal{N}$ and $\mathcal{S}$ at zero temperature $T=0$
\begin{equation}
	\label{eq3}
	\mathcal{N}(T=0) = \sqrt{1 + 4 \mathcal{S} - 8 \mathcal{S}^{2}} - 2 \mathcal{S}
\end{equation}
for general (A)LM phases having the SU(2) symmetry.
The perfect screening yields $\mathcal{S} = 0$ and $\mathcal{N} = 1$, while no screening implies $\mathcal{S} = 1/2$ and $\mathcal{N} = 0$. $\mathcal{N}$ is larger for more screening.
By contrast, the Kondo phase obeys
another connection $\mathcal{N}(T=0) = \sqrt{1-4\mathcal{S}^{2}}$~\cite{Kim21}.
The distinction arises, since the LM and Kondo phases correspond to the different fixed points having two-fold degenerate and non-degenerate ground states, respectively.
It is interesting to have such relations between the entanglement and a local observable at the fixed points.
 
\begin{figure}[t]
	\centerline{\includegraphics[scale=0.69]{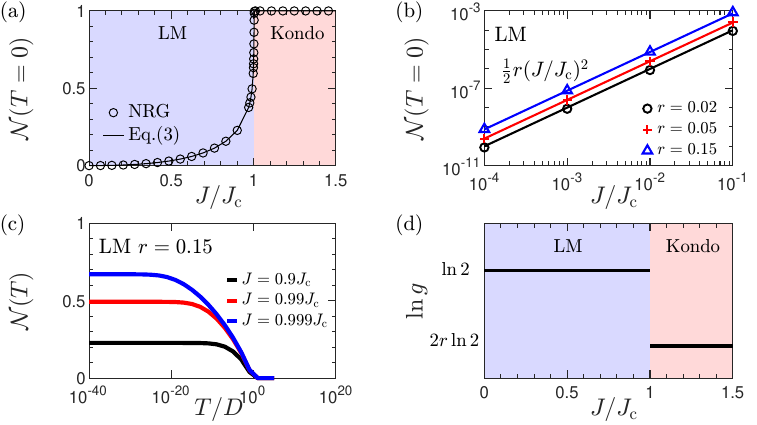}}
	\caption{Pseudogap DOS.
		(a) Zero-temperature entanglement negativity $\mathcal{N}$ as a function of $J$.
		It is directly computed (circles) by NRG~\cite{Shim18} and indirectly (curves) by combining the NRG results of $\mathcal{S}$ and Eq.~\eqref{eq3}. $r=0.15$ and $J_{\text{c}}/D\approx0.29$ are chosen.  
		(b) NRG results (markers) of $\mathcal{N}(J)$ at $J \ll J_\text{c}$ and zero temperature in the LM phase with various $r$'s. They agree with Eq.~\eqref{eq4} (lines).
		(c) NRG results of $\mathcal{N}(T)$ at finite temperature $T$ in the LM phase with various $J$'s. 
		(d) Impurity entropy $\ln g$ at zero temperature as a function of $J$.
	}
	\label{fig1}
\end{figure}

We compute $\mathcal{N}(T=0)$ for the LM and Kondo phases of the pseudogap DOS by using NRG~\cite{Wilson75, Bulla08, Shim18, Supp}.
In the LM phase, we find $0 < \mathcal{N} < 1$ for nonzero $J$
and derive $\mathcal{N}$ for small $J\ll J_{\mathrm{c}}$ and $r \ll 1$,
\begin{equation}
	\label{eq4}
	\mathcal{N}(T=0) = \frac{1}{2}r\left(\frac{J}{J_\mathrm{c}}\right)^2 + O\big((J/J_{\text{c}})^{3}\big),
\end{equation}
using perturbative renormalization group (RG) methods~\cite{Moca21, Withoff90, Supp}.
The NRG results agree with Eq.~\eqref{eq3} [Fig.~\ref{fig1}(a)] and Eq.~\eqref{eq4} [Fig.~\ref{fig1}(b)],
This shows that the impurity spin is screened, albeit not perfectly, in the LM phase.
More screening happens at larger $J$.
By contrast, the Kondo phase shows the maximum value~\cite{Kim21} of $\mathcal{N} = 1$ independent of $J$, implying the perfect spin screening.

This shows that the LM phase at zero temperature differs from the high-temperature phase (also called a ``local moment'' phase) of usual Kondo effects where no screening happens and the entanglement vanishes~\cite{Kim21,Shim23}.

Figure~\ref{fig1}(c) shows thermal suppression of the spin screening and the entanglement $\mathcal{N}$ in the LM phase. Their low-temperature suppression is discussed later.

Certain thermodynamic quantities such as the impurity entropy $\ln g$ are unsuitable to see the spin screening, contrary to common wisdom.
At zero temperature, $\ln g = \ln 2$ regardless of $J$ in the LM phase [Fig.~\ref{fig1}(d)]~\cite{Buxton98}.
The value $\ln 2$ means that the impurity contributes to the free energy like a decoupled spin-1/2 \cite{Vojta06, Wagner18}.
This should not be interpreted such that the impurity is a free spin not screened by conduction electrons.
Moreover, while in usual Kondo effects $\ln g = 0$ implies the perfect screening,
another value of $\ln g = 2r \ln 2$ is found~\cite{Buxton98} in the Kondo phase of the pseudogap DOS, which also exhibits the perfect screening. 
The impurity entropy, indicating ``spin degeneracy''~\cite{Affleck91,Vojta06}, can not detect the screening.

The entanglement negativity, applicable to mixed states, is suitable for quantifying the spin screening in LM phases, which are in mixed states not only at finite temperatures but also at zero temperature~\cite{LMdegeneracy}.
We note that the entanglement entropy~\cite{Pixley15, Wagner18} is improper for the thermodynamic ground state of LM phases, as it is an entanglement measure valid only for pure states.

\begin{figure}[t]
	\includegraphics[scale=0.69]{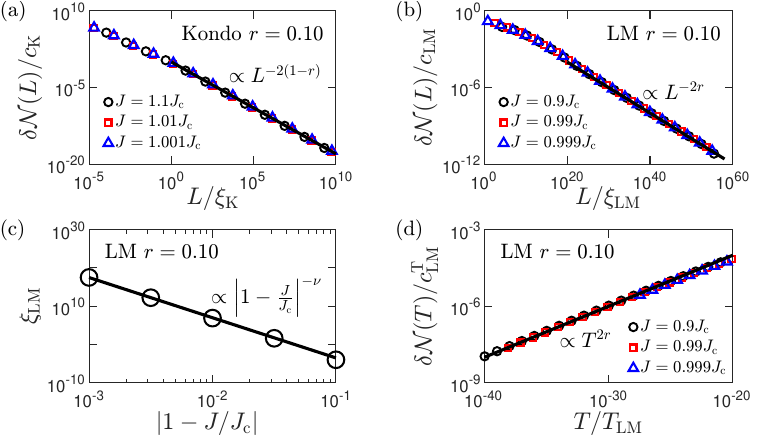}
	\caption
	{ NRG results (markers) of spin clouds for the pseudogap DOS.
		(a,b) Spatial distribution $\delta \mathcal{N}(L)$ of the spin cloud in the (a) Kondo and (b) LM phases.
		Its tail (at $L \gg \xi_\text{K}$ or $\xi_\text{LM}$) follows the power law in Eq.~\eqref{eq5} with the exponent $\alpha$ in Table~\ref{table1}. We choose $r=0.1$ and $J_{\text{c}}/D\approx 0.19$.
		(c) In the LM phase, the cloud length $\xi_{\text{LM}}$ is proportional to $|1-J/J_{\text{c}}|^{-\nu}$.
		The NRG results show $\nu=10.5$ at $r=0.1$. 
		(d) Thermal suppression of the entanglement in the LM phase. It follows the power law in Eq.~\eqref{eq6} at $T \ll T_\text{LM}$.
	}
	\label{fig2}
\end{figure}

{\it Spin cloud for the pseudogap DOS.---}
We show that the ``LM spin screening cloud'', analogous to the Kondo cloud, is formed in the pseudogap LM phase. To quantify its spatial distribution, we apply a local perturbation~\cite{LSB} to electrons at distance $L$ from the impurity, which breaks a (either SU(2) or particle-hole) symmetry of the system~\cite{Supp}.
This local symmetry breaking (LSB) makes the entanglement deviate from the value $\mathcal{N}$ in the absence of the LSB
to $\mathcal{N}'(L)$ in the presence at $L$.
The deviation $\delta \mathcal{N}(L) \equiv |\mathcal{N} - \mathcal{N}'(L)|$ quantifies the spatial distribution of the cloud, as more deviation means more contribution to the cloud at the distance $L$~\cite{Shim23}.

\newcolumntype{C}{>{\centering\arraybackslash}p{0.11\textwidth}}
\begin{table}[b]
	\begin{center}
		\begin{tabular}{C C C C} 
			\hline\hline
			$\alpha$ & $0 < |r| < 0.5$ & $0.5 \leq |r| < 1$ & $1 \leq |r|$ \\ 
			\hline
			Kondo & $2(1-r)$ & --- & --- \\
			LM & $2r$ & $2r$ & $1+r$ \\
			ALM & $-2r$ & $-2r$ & --- \\
			\hline\hline
		\end{tabular}
		\caption{
			Exponent $\alpha$ in Eqs.~\eqref{eq5} and~\eqref{eq6} for the Kondo and LM phases of the pseudogap DOS and the ALM phase of the diverging DOS. The blanks mean the absence of phases.
		}
		\label{table1}
	\end{center}
\end{table}

The spin cloud appears in both the Kondo and LM phases of the pseudogap DOS.
We derive their power-law spatial decay at large $L \gg \xi_\text{K}$ or $\xi_\text{LM}$, 
\begin{equation}
	\label{eq5}
	\delta\mathcal{N}(L) =
	\begin{cases}
		c_{\text{K}}(L/\xi_{\text{K}})^{-\alpha_\text{K}} & \text{Kondo}, \\
		c_{\text{LM}}(L/\xi_{\text{LM}})^{-\alpha_\text{LM}} & \text{LM},
	\end{cases}
\end{equation}
combining Eq.~\eqref{eq3} and our fixed-point analysis~\cite{Wilson75,Buxton98,Supp} of $\mathcal{S}$ (discussed later). The exponent $\alpha_\text{K,LM}$ found in Table~\ref{table1} agrees with NRG results.
The power-law decay of the cloud in the Kondo phase [Fig.~\ref{fig2} (a)] differs from that of the conventional Kondo cloud whose exponent is $\alpha = 2$~\cite{decaystozero,Affleck09,Shim23}.
The LM cloud also exhibits a power-law decay but with another exponent [Fig.~\ref{fig2} (b)].
The prefactor $c_{\text{K}}$ is independent of $L$ and $J$ in the Kondo phase, while
$c_{\text{LM}}$ is independent of $L$ but depends on $|1-J/J_{\text{c}}|$ in the LM phase~\cite{Supp}.
The power-law decay is universal in both phases, as $\delta\mathcal{N}(L)$ lies on a single curve as a function of the ratio of $L$ to the Kondo $\xi_{\text{K}}$ or LM cloud length $\xi_{\text{LM}}$ for different $J$'s.

At $r \ll 1$, we find~\cite{Supp} that 
the Kondo and LM cloud lengths emerge as scaling invariant quantities in the RG flow~\cite{Hewson97,Withoff90}, following the same scaling of $\xi_{\text{LM, K}}\propto|1-J/J_{\text{c}}|^{-1/r}$. The emergence of the finite length is interesting in the LM phase where the RG flow implies vanishing $J$ at low energies.
At the other $r$'s, our NRG results [Fig.~\ref{fig2}(c) and Ref.~\cite{Supp}] show that for each $r$ they still follow the same, but modified, scaling of $\xi_{\text{LM, K}}\propto|1-J/J_{\text{c}}|^{-\nu}$ with exponent $\nu$ non-trivially depending on $r$.
This is consistent with the scaling hypothesis near the quantum critical point~\cite{Vojta06}.

The universality is also found in other quantities. The change $\delta \mathcal{S} (L) \equiv |\mathcal{S} - \mathcal{S}' (L)|$ of the ground-state impurity spin $\mathcal{S} = |\langle\Psi_{\text{GS}}|\vec{S}_{\mathrm{imp}}|\Psi_{\text{GS}}\rangle|$ by the LSB exhibits a universal power law $\delta S(L) = c (L / \xi)^{- \alpha_\mathcal{S}}$ at $L \gg \xi_\text{K}$ or $\xi_\textrm{LM}$, where
$\mathcal{S}'$ is the impurity spin value in the presence of the LSB at $L$~\cite{Supp}.
Here $\xi = \xi_\text{K}$ and $2\alpha_\mathcal{S} = \alpha_\text{K}$ ($\xi = \xi_\text{LM}$, $\alpha_\mathcal{S} = \alpha_\text{LM}$) in the Kondo (LM) phase. $c$ is a constant independent of $J$ and $L$ in both phases. This supports the appearance of the spin clouds. We note that $\delta \mathcal{N} (L) \propto \mathcal{S} \times \delta \mathcal{S} (L)$ in the LM phase, according to Eq.~\eqref{eq3}, while $\delta \mathcal{N} (L) \propto \big( \delta \mathcal{S} (L) \big)^2$ in the Kondo phase~\cite{Shim23}.
 
The thermal suppression $\mathcal{N}(T)$ of the spin cloud at temperature $T 
\ll T_{\text{K}}$ or $T_{\text{LM}}$ also follows a universal power law in both the Kondo and LM phases,
\begin{equation}
	\label{eq6}
	\delta\mathcal{N}(T) =
	\begin{cases}
	c_{\text{K}}^\text{T} (T/T_{\text{K}})^{\alpha_\text{K}} & \text{Kondo}, \\
	c_{\text{LM}}^\text{T} (T/T_{\text{LM}})^{\alpha_\text{LM}} & \text{LM},
	\end{cases}
\end{equation}
where $\delta \mathcal{N} (T) \equiv | \mathcal{N}(T=0) - \mathcal{N}(T)|$ [Fig.~\ref{fig2}(d) for the LM phase].
The LM temperature $T_{\text{LM}} \propto 1/\xi_{\text{LM}}$ corresponds to the Kondo temperature $T_{\text{K}} \propto 1/\xi_{\text{K}}$ in the Kondo phase. The power-law exponents are the same with Eq.~\eqref{eq5} and Table~\ref{table1}.
While $c_{\text{K}}^\text{T}$ is independent of $T$ and $J$, 
$c_{\text{LM}}^\text{T}$ is independent of $T$ but depends on $|1-J/J_{\text{c}}|$.

\begin{figure}[t]
	\centerline{\includegraphics[scale=0.69]{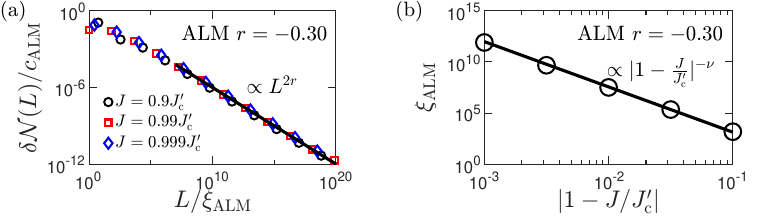}}
	\caption
	{		
		NRG results of the spatial decay of spin clouds for the diverging DOS with $r=-0.3$. We use $V/D=0.4$ and $J'_{\text{c}}/D\approx0.32$.
		(a) The decay follows the universal power law of the ALM phase.
		(b) The cloud length follows $\xi_{\text{ALM}}\propto |1-J/J'_{\text{c}}|^{-\nu}$. The NRG results show $\nu=4.4$.
	}
	\label{fig3}
\end{figure}

{\it Spin cloud for the diverging DOS.---}
In the ALM phase for the diverging DOS, universal spin clouds also appear.
They follow the power law in Eq.~\eqref{eq5} with exponent $\alpha_\text{ALM}$ [Table~\ref{table1}]
and prefactor $c_{\text{ALM}}$ that depends on $|1-J/J'_{\text{c}}|$.
The cloud length follows $\xi_{\text{ALM}}\propto |1-J/J'_{\text{c}}|^{-\nu}$. The exponent $\nu$ non-trivially depends on $r$ and the potential scattering strength $V$. 
These agree with NRG results [Fig.~\ref{fig3}].

{\it Fixed point analysis.---} We derive all the above power-law spatial decay and thermal suppression of the LM clouds by developing the fixed point analysis~\cite{Wilson75,Buxton98}. It utilizes the finite NRG Wilson chain whose energy scale (inverse energy) corresponds to the temperature (the distance $L$ between the impurity and the LSB). The finite chain is described by the fixed-point Hamiltonian for the target phase and possible perturbation terms by irrelevant operators preserving the symmetries of the phase.
The power laws are obtained~\cite{Supp} by constructing the perturbed eigenstates, computing the chain-length dependence of $\mathcal{S}$, and using Eq.~\eqref{eq3}. This is a powerful approach for phases where standard methods such as conformal field theories are inapplicable.

\begin{figure}[t]
	\centerline{\includegraphics[scale=0.69]{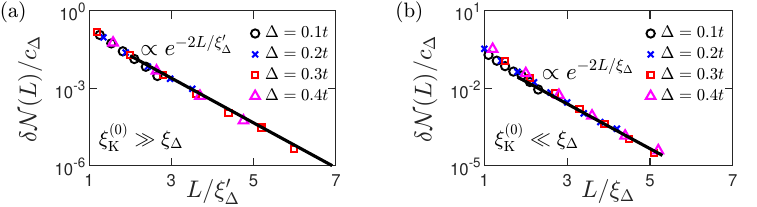}}
	\caption
	{
		DMRG results of the spatial decay of LM clouds for the hard gap DOS when (a) $\xi_{\text{K}}^{(0)} \gg \xi_{\Delta}$ and (b) $\xi_{\text{K}}^{(0)} \ll \xi_{\Delta}$.
		It follows the universal exponential decay in Eq.~\eqref{eq7}.
		The SSH lattice having 200 sites is used to describe an electron bath of the hard gap $\Delta$.
		In (a), the cloud length $\xi_\text{LM}$ is found as $\xi_{\Delta}'\approx 1.40, 1.35, 1.33, 1.32 v_{\mathrm{F}}/\Delta$ for $\Delta = 0.1, 0.2, 0.3, 0.4t$, respectively. We choose $J=0.10t$ and $\xi_{\text{K}}^{(0)}>10^{5}$ sites, where $t$ is the average hopping energy of the SSH model. In (b), we find $\xi_\text{LM} = \xi_\Delta$. $J=16t$ and $\xi_{\text{K}}^{(0)}< 0.03$ sites. 
	}
	\label{fig4}
\end{figure}

{\it Spin cloud for the hard gap DOS.---}
For the hard gap DOS, a spin cloud also emerges.
Figure~\eqref{fig4} shows DMRG results of the impurity spin coupled with the SSH lattice with the hard gap $\Delta$, where the LSB is applied (see the details in Ref.~\cite{Supp}).
The spatial decay of the cloud follows a universal exponential decay 
\begin{equation}
	\label{eq7}
	\delta\mathcal{N}(L) \propto c_{\Delta}\exp(-2L/\xi_{\text{LM}})
\end{equation}
with constant $c_{\Delta}$ due to the gap $\Delta$.
The LM cloud length $\xi_{\text{LM}}$ is determined by competition between the bare Kondo cloud length~\cite{Sorensen96} $\xi_{\text{K}}^{(0)}$ in the absence of the gap and the gap correlation length~\cite{Moca21} $\xi_{\Delta}=v_{\mathrm{F}}/\Delta$ (with Fermi velocity $v_{\mathrm{F}}$).
When $\xi_{\text{K}}^{(0)}\ll \xi_{\Delta}$, the LM cloud length follows $\xi_{\text{LM}}=\xi_{\Delta}$, as found with superconducting baths~\cite{Moca21}, and $c_\Delta \propto \Delta^3$.
When $\xi_{\text{K}}^{(0)}\gg \xi_{\Delta}$, on the other hand, the LM length equals another length $\xi_{\text{LM}}=\xi_{\Delta}'$, which is determined by not only $\Delta$ but also the detailed DOS form, and $c_\Delta$ is independent of $\Delta$.
The DMRG results and Eq.~\eqref{eq7} agree with the perturbation theory~\cite{Supp} at $\xi_{\text{K}}^{(0)}\gg \xi_{\Delta}$ where $J$ is a perturbation parameter.
 
{\it Discussion.---}
We have shown that a magnetic impurity is screened, albeit partially, by conduction electrons in all possible LM phases. They form a spin cloud, whose spatial distribution and thermal suppression follow universal scaling (power-law or exponential) functions characterized by the LM cloud length or LM temperature. 

In the LM phases, the impurity spin decouples from conduction electrons at the fixed point in the context of the RG flow. The spin cloud is not described by this standard long-range or low-energy perspective of fixed points. Instead, we predict that there is the finite length or energy scaling-invariant under the RG flow, the LM cloud length or LM temperature. This guarantees the emergence of the spin cloud at nonvanishing coupling $J$.
 
The LM cloud generalizes the Kondo cloud. However it has fundamental difference as the LM and Kondo phases correspond to the different fixed points having two-fold degenerate and non-degenerate ground states, respectively. The LM cloud partially screens the impurity, having different scaling behaviors from the Kondo cloud.
It is identified by the negativity, a mixed-state entanglement measure. It cannot be detected by the entanglement entropy even at zero temperature
nor by thermodynamic quantities such as the impurity entropy.

Our work offers a unified framework applicable to diverse systems.
For example, graphene provides pseudogap DOS with $r=1$ when the Fermi level lies at the Dirac point.
Our NRG calculation (not shown) demonstrates that an LM cloud is formed around 
a magnetic impurity in graphene. 
It follows $\delta\mathcal{N}(L)\propto L^{-2}$ in agreement with Table~\ref{table1}.
It will be interesting to experimentally detect LM clouds in ways similar to Kondo cloud detection~\cite{Borzenets20} or utilizing a relation~\cite{Yoo18} between the entanglement and a local observable like Eq.~\eqref{eq3}.

Our study provides fundamental understanding of how spin entanglement forms and spreads over space in general impurity systems.
It is valuable to apply our findings to pseudogap Kondo systems in strongly correlated environments like heavy fermion systems and the Kondo destruction quantum critical point~\cite{Si01,Schroder00}. 
It is interesting to see whether a cloud appears in nontrivial situations,
for example, two magnetic impurities in a pseudogap system (cf. two-channel or two-impurity Kondo clouds~\cite{Shim23}).
Our fixed point analysis will be useful for those systems.

We thank Matthias Vojta and S.-S. B. Lee for useful comments. This work is supported by Korea NRF (grant number 2023R1A2C2003430; SRC Center for Quantum Coherence in Condensed Matter, grant number RS-2023-00207732). D.K. acknowledges support by Korea NRF via Basic Science Research Program for Ph.D. students (2022R1A6A3A13062095).

%%%%%%%%%%%%%%   Supplementary Material   %%%%%%%%%%%%%

\clearpage
\newpage

\onecolumngrid

\begin{center}
\textbf{\large Supplementary Material for \textquotedblleft Universal Spin Screening Clouds in Local Moment Phases \textquotedblright}
\end{center}

\begin{center}
Minsoo L. Kim,\textsuperscript{1} Jeongmin Shim,\textsuperscript{1} H.-S. Sim,\textsuperscript{1,*} and Donghoon Kim\textsuperscript{1,$\dagger$}

\textit{\small \textsuperscript{1}Department of Physics, Korea Advanced Institute of Science and Technology, Daejeon 34141, Korea}

\end{center}

\newcommand{\beginsup}{
        \setcounter{equation}{0}
        \renewcommand{\theequation}{S\arabic{equation}}
        \setcounter{table}{0}
        \renewcommand{\thetable}{S\arabic{table}}
        \setcounter{figure}{0}
        \renewcommand{\thefigure}{S\arabic{figure}}
     }
\beginsup

This material contains the derivation of Eqs.~(3)-(6) of the main text, the scaling arguments for the LM cloud length, the fixed point analysis for the power-law behaviors in (A)LM phases, the details of the NRG calculations, NRG supports of the universality of the spin clouds, and the DMRG calculations and the perturbation theory for the SSH model.
For simplicity, we below set $\hbar\equiv1$, $k_{\text{B}}\equiv 1$, and the Fermi velocity $v_{\mathrm{F}}\equiv1$.

\section{I. Entanglement and spin cloud for pseudogap DOS: Analytic results}
\label{sec:I}
\subsection{Derivation of Eq.~(3): Ground-state entanglement in the local moment phase}
We derive Eq.~(3) of the main text, the entanglement negativity formula at zero temperature in the local moment (LM) phase. We consider the following Hamiltonian,
\begin{align}
	\label{KondoHam}
	H_{\text{K}}
	=\sum_{\mathbf{k},\sigma} {\epsilon}_{\mathbf{k}}\,{c_{\mathbf{k}\sigma}^\dagger}{c_{\mathbf{k}\sigma}}+J\vec{S}_{\text{imp}}\cdot\vec{S}_{\text{bath}}
	+ V \sum_{\mathbf{k}, \mathbf{k}', \sigma} c_{\mathbf{k}\sigma}^\dagger c_{\mathbf{k}'\sigma}.
\end{align}
Here $c_{\mathbf{k}\sigma}$ is the annihilation operator of a bath conduction electron with the momentum $\mathbf{k}$ and the spin $\sigma$, $J$ is the Kondo coupling, $\vec{S}_{\text{bath}} = \sum_{\sigma,\sigma'}\sum_{\mathbf{k},\mathbf{k}'}c_{\mathbf{k}\sigma}^{\dagger}\frac{\vec{\sigma}_{\sigma,\sigma'}}{2}c_{\mathbf{k}'\sigma'}$ is the spin operator of bath electrons at the impurity position,  $\vec{\sigma}$ is the Pauli matrix vector, $\vec{S}_{\mathrm{imp}}$ is the impurity spin, and $V$ is a potential scattering.
We consider arbitrary energy-dependent density of states (DOS) $\rho(\epsilon)$ of the electron bath at the impurity position.

First, we express the ground states of the Hamiltonian in the bipartite form where the whole system is partitioned into the impurity and the bath conduction electrons.
The Hamiltonian $H_{\text{K}}$ commutes with the total spin operator $\vec{S}_{\text{tot}} = \vec{S}_{\text{imp}} + \vec{S}_{\text{bath}}$ so that each energy eigenstate $|E \rangle$ has the spin quantum numbers $(j_{\text{tot}},j_{\text{tot}}^{z})$ satisfying $\vec{S}_{\text{tot}}^{2} |E \rangle = j_{\text{tot}}(j_{\text{tot}} + 1) |E \rangle$ and $S_{\text{tot}}^{z} |E \rangle = j_{\text{tot}}^{z} |E \rangle$.
The ground states in the local moment phase are two-fold degenerate with $(j_{\mathrm{tot}},j_{\mathrm{tot}}^{z}) = (1/2,\pm 1/2)$ because they have the same quantum number with the ground states in the case that the Kondo coupling vanishes. We denote these ground states as
\begin{align}\label{twoGSQN}
|\Psi_{\mathrm{GS},+} \rangle = |j_{\mathrm{tot}} = 1/2,j_{\mathrm{tot}}^{z} = 1/2 \rangle, \qquad |\Psi_{\mathrm{GS},-} \rangle = |j_{\mathrm{tot}} = 1/2,j_{\mathrm{tot}}^{z} = -1/2 \rangle,
\end{align}
and introduce $\mathcal{S}^z \equiv \langle \Psi_{\mathrm{GS},+} | S_{\mathrm{imp}}^{z} | \Psi_{\mathrm{GS},+} \rangle$, the expectation value of the impurity spin operator $S_{\mathrm{imp}}^{z}$ in the state $|\Psi_{\mathrm{GS},+} \rangle$.
$\mathcal{S}^{z}$ is positive, since $| \Psi_{\mathrm{GS},+} \rangle$ is an eigenstate of $S_{\mathrm{tot}}^{z}$ with $j_{\mathrm{tot}}^{z}=+1/2$.
The relation $|\Psi_{\mathrm{GS},-} \rangle \propto e^{-i \pi S_{\mathrm{tot}}^{x}} |\Psi_{\mathrm{GS},+} \rangle$ by the $\mathrm{SU}(2)$ symmetry yields
\begin{align}
\langle \Psi_{\mathrm{GS},-} | S_{\mathrm{imp}}^{z} | \Psi_{\mathrm{GS},-} \rangle = \langle \Psi_{\mathrm{GS},+} | e^{i \pi S_{\mathrm{tot}}^{x}} S_{\mathrm{imp}}^{z} e^{-i \pi S_{\mathrm{tot}}^{x}} |\Psi_{\mathrm{GS},+} \rangle = - \langle \Psi_{\mathrm{GS},+} | S_{\mathrm{imp}}^{z} | \Psi_{\mathrm{GS},+} \rangle = - \mathcal{S}^z.
\end{align}
Since $\langle \Psi_{\mathrm{GS},+} | S_{\mathrm{imp}}^{x} | \Psi_{\mathrm{GS},+} \rangle = \langle \Psi_{\mathrm{GS},+} | S_{\mathrm{imp}}^{y} | \Psi_{\mathrm{GS},+} \rangle = 0$, the quantity $\mathcal{S} \equiv |\langle \Psi_\text{GS} | \vec{S}_\text{imp} | \Psi_\text{GS} \rangle|$ introduced in the main text equals $\mathcal{S}^z$ (this equivalence will be established in the final part of the derivation). 
For later use, we set the phase factor to be $i$, i.e., $|\Psi_{\mathrm{GS},-} \rangle = i e^{-i \pi S_{\mathrm{tot}}^{x}} |\Psi_{\mathrm{GS},+} \rangle$.

We represent the ground states in terms of the tensor product of two irreducible representations of the impurity and the bath. Here we divide the quantum numbers $(j_{\mathrm{tot}},j_{\mathrm{tot}}^{z})$ into those of the impurity and the bath, $(j_{\mathrm{imp}},j_{\mathrm{imp}}^{z})$ and $(j_{\mathrm{bath}},j_{\mathrm{bath}}^{z})$, satisfying
$\vec{S}_{\text{s}}^{2} |j_{\text{s}},j_{\text{s}}^{z} \rangle = j_{\text{s}}(j_{\text{s}} + 1) |j_{\text{s}},j_{\text{s}}^{z} \rangle$ and $S_{\text{s}}^{z} |j_{\text{s}},j_{\text{s}}^{z} \rangle = j_{\text{s}}^{z} |j_{\text{s}},j_{\text{s}}^{z} \rangle$ for $\text{s} = \text{imp},\text{bath}$. 
The ground states are written, according to the summation of the angular momentum and the Clebsch-Gordan coefficients, as
\begin{align}
	\label{psiGSplus}
	|\Psi_{\mathrm{GS},+} \rangle &= a \Bigg(\sqrt{\frac{2}{3}} |{\downarrow} \rangle |{+1} \rangle - \sqrt{\frac{1}{3}} |{\uparrow} \rangle |{0} \rangle\Bigg) + b |{\uparrow} \rangle |{\phi} \rangle, \\
	\label{psiGSminus}
	|\Psi_{\mathrm{GS},-} \rangle &= c \Bigg(\sqrt{\frac{1}{3}} |{\downarrow} \rangle |{0} \rangle - \sqrt{\frac{2}{3}} |{\uparrow} \rangle |{-1} \rangle\Bigg) + d |{\downarrow} \rangle |{\phi} \rangle,
\end{align}
with the coefficients $a,b,c,d \in \mathbb{C}$ satisfying the normalization condition $|a|^{2} + |b|^{2} = |c|^{2} + |d|^{2} = 1$, where $|{\uparrow} \rangle$ and $|{\downarrow} \rangle$ are the impurity states with the spin quantum number $(j_{\mathrm{imp}},j_{\mathrm{imp}}^{z}) = (1/2,1/2)$ and $(1/2,-1/2)$, and 
$|{+1} \rangle$, $|{0} \rangle$, $|{-1} \rangle$, and $|{\phi} \rangle$ are the bath states with the spin quantum number $(j_{\mathrm{bath}},j_{\mathrm{bath}}^{z}) = (1,1)$, $(1,0)$, $(1,-1)$, and $(0,0)$ respectively.
We combine the normalization condition with the relation $\langle \Psi_{\mathrm{GS},\pm}| S_{\mathrm{imp}}^{z} |\Psi_{\mathrm{GS},\pm} \rangle = \pm \mathcal{S}$, which is equivalent to $|b|^{2} - \frac{1}{3} |a|^{2} = |d|^{2} - \frac{1}{3} |c|^{2} = 2 \mathcal{S}$, to obtain  
\begin{align}\label{coefficients}
|a|^{2} = \frac{3 - 6 \mathcal{S}}{4}, \qquad |b|^{2} = \frac{1 + 6 \mathcal{S}}{4}, \qquad |c|^{2} = \frac{3 - 6 \mathcal{S}}{4}, \qquad |d|^{2} = \frac{1 + 6 \mathcal{S}}{4}.
\end{align}
We derive one more identity for $a,b,c,$ and $d$ by using the relation of $S_{\mathrm{tot}}^{x} |\Psi_{\mathrm{GS},+} \rangle=\frac{1}{2} |\Psi_{\mathrm{GS},-} \rangle$,
\begin{align}\label{Stotxidentity}
\frac{1}{2} \Bigg[a \Bigg(\sqrt{\frac{1}{3}} |{\downarrow} \rangle |{0} \rangle - \sqrt{\frac{2}{3}} |{\uparrow} \rangle |{-1} \rangle\Bigg) + b |{\downarrow} \rangle |{\phi} \rangle\Bigg] = \frac{1}{2} \Bigg[c\Bigg(\sqrt{\frac{1}{3}} |{\downarrow} \rangle|0 \rangle - \sqrt{\frac{2}{3}} |{\uparrow} \rangle |{-1} \rangle\Bigg) + d |{\downarrow} \rangle |{\phi} \rangle\Bigg].
\end{align}
Combining the equations, we find (up to an overall phase factor)
\begin{align}
a = \sqrt{\frac{3 - 6 \mathcal{S}}{4}}, \qquad b = e^{i \alpha} \sqrt{\frac{1 + 6 \mathcal{S}}{4}}, \qquad c = \sqrt{\frac{3 - 6 \mathcal{S}}{4}}, \qquad d = e^{i {\varphi}} \sqrt{\frac{1 + 6 \mathcal{S}}{4}},
\end{align}
with $\varphi \in \mathbb{R}$. As the phase factor $e^{i \varphi}$ is absorbed into $|\phi \rangle$ ($|\phi \rangle \rightarrow e^{i \varphi} |\phi \rangle$), we get the expression of the ground states,
\begin{align}
|\Psi_{\mathrm{GS},+} \rangle &= \sqrt{\frac{1 - 2 \mathcal{S}}{2}} |{\downarrow} \rangle |{+1} \rangle - \sqrt{\frac{1 - 2 \mathcal{S}}{4}} |{\uparrow} \rangle |{0} \rangle + \sqrt{\frac{1 + 6 \mathcal{S}}{4}} |{\uparrow} \rangle |{\phi} \rangle, \label{GSlocalmoment1} \\
|\Psi_{\mathrm{GS},-} \rangle &= \sqrt{\frac{1 - 2 \mathcal{S}}{4}} |{\downarrow} \rangle |{0} \rangle - \sqrt{\frac{1 - 2 \mathcal{S}}{2}} |{\uparrow} \rangle |{-1} \rangle + \sqrt{\frac{1 + 6 \mathcal{S}}{4}} |{\downarrow} \rangle |{\phi} \rangle. \label{GSlocalmoment2}
\end{align}

Using Eqs.~\eqref{GSlocalmoment1} \eqref{GSlocalmoment2}, we construct
the density matrix in the ground state manifold, $\hat{\rho}_\text{GS} = \frac{1}{2} |\Psi_{\mathrm{GS},+} \rangle \langle \Psi_{\mathrm{GS},+}| + \frac{1}{2} |\Psi_{\mathrm{GS},-} \rangle \langle \Psi_{\mathrm{GS},-}|$, which is a $8 \times 8$ matrix in the basis $\{|{\uparrow} \rangle |{+1} \rangle, |{\uparrow} \rangle |{0} \rangle, |{\uparrow} \rangle |{-1} \rangle,|{\uparrow} \rangle |{\phi} \rangle,|{\downarrow} \rangle |{+1} \rangle, |{\downarrow} \rangle |{0} \rangle, |{\downarrow} \rangle |{-1} \rangle,|{\downarrow} \rangle |{\phi} \rangle\}$.
The partial transpose ($\mathrm{T}_{\mathrm{imp}}$) of $\hat{\rho}_\text{GS}$ with respect to the impurity in this basis becomes
\begin{align}
\hat{\rho}_\text{GS}^{\mathrm{T}_{\text{imp}}} \doteq \begin{pmatrix}
0 & 0 & 0 & 0 & 0 & -\frac{1-2 \mathcal{S}}{4 \sqrt{2}} & 0 & \frac{\sqrt{1-2 \mathcal{S}} \sqrt{1 + 6 \mathcal{S}}}{4\sqrt{2}} \\
0 & \frac{1 - 2 \mathcal{S}}{8} & 0 & - \frac{\sqrt{1 - 2 \mathcal{S}} \sqrt{1 + 6 \mathcal{S}}}{8} & 0 & 0 & - \frac{1 - 2 \mathcal{S}}{4 \sqrt{2}} & 0 \\
0 & 0 & \frac{1 - 2 \mathcal{S}}{4} & 0 & 0 & 0 & 0 & 0 \\
0 & - \frac{\sqrt{1 - 2 \mathcal{S}} \sqrt{1 + 6 \mathcal{S}}}{8} & 0 & \frac{1 + 6 \mathcal{S}}{8} & 0 & 0 & - \frac{\sqrt{1 - 2 \mathcal{S}} \sqrt{1 + 6 \mathcal{S}}}{4 \sqrt{2}} & 0 \\
0 & 0 & 0 & 0 & \frac{1 - 2 \mathcal{S}}{4} & 0 & 0 & 0 \\
- \frac{1 - 2 \mathcal{S}}{4 \sqrt{2}} & 0 & 0 & 0 & 0 & \frac{1 - 2 \mathcal{S}}{8} & 0 & \frac{\sqrt{1 - 2 \mathcal{S}} \sqrt{1 + 6 \mathcal{S}}}{8} \\
0 & - \frac{1 - 2 \mathcal{S}}{4 \sqrt{2}} & 0 & - \frac{\sqrt{1 - 2 \mathcal{S}} \sqrt{1 + 6 \mathcal{S}}}{4 \sqrt{2}} & 0 & 0 & 0 & 0 \\
\frac{\sqrt{1 - 2 \mathcal{S}} \sqrt{1 + 6 \mathcal{S}}}{4 \sqrt{2}} & 0 & 0 & 0 & 0 & \frac{\sqrt{1 - 2 \mathcal{S}} \sqrt{1 + 6 \mathcal{S}}}{8} & 0 & \frac{1 + 6 \mathcal{S}}{8}
\end{pmatrix}. \nonumber
\end{align}
Note that $0 \leq \mathcal{S} \leq 1/2$.
The singular values of $\hat{\rho}_\text{GS}^{\mathrm{T}_{\text{imp}}}$ are four $\frac{1 - 2 \mathcal{S}}{4}$, two $\frac{\sqrt{1 + 4 \mathcal{S} - 8 \mathcal{S}^{2}} - 2\mathcal{S}}{4}$, and two $\frac{\sqrt{1 + 4 \mathcal{S} - 8 \mathcal{S}^{2}} + 2\mathcal{S}}{4}$.
The entanglement negativity $\mathcal{N}$ between the impurity and the bath at zero temperature is found as
\begin{align}
\label{LMNformula}
\mathcal{N}(T=0) = \Vert \hat{\rho}_\text{GS}^{\mathrm{T}_{\text{imp}}} \Vert_{1} - 1 = \sqrt{1 + 4 \mathcal{S} - 8 \mathcal{S}^{2}} - 2 \mathcal{S},
\end{align}
where $\Vert \rho_\text{GS}^{\mathrm{T}_{\mathrm{imp}}} \Vert_{1}$ is the trace norm of $\hat{\rho}_\text{GS}^{\mathrm{T}_{\text{imp}}}$ that equals the sum of all the singular values of $\hat{\rho}_\text{GS}^{\mathrm{T}_{\text{imp}}}$. 
For the ground states $|\Psi_\text{GS, $\pm$} \rangle$,
Eq.~\eqref{LMNformula} provides Eq.~(3) of the main text.

Next, we prove Eq.~(3) of the main text for any arbitrary superposition $|\psi \rangle$ of $|\Psi_\text{GS, $+$} \rangle$ and $|\Psi_\text{GS, $-$} \rangle$. We represent the superposition as $|\psi \rangle = e^{- i \theta \vec{n} \cdot \vec{S}_{\mathrm{tot}}} |\Psi_{\mathrm{GS},+} \rangle$ where $\theta \in [0, 2\pi)$ and $\vec{n}$ is a unit vector in three-dimensional space.
Given that $\vec{S}_{\mathrm{tot}} = \vec{S}_{\mathrm{imp}} + \vec{S}_{\mathrm{bath}}$ and the commutator $[S_{\mathrm{imp}}^{\alpha}, S_{\mathrm{bath}}^{\beta}] = 0$ for all \(\alpha, \beta = x, y, z\), we find the expectation value of
\begin{align}
	\langle \psi | \vec{S}_{\mathrm{imp}} | \psi \rangle = \langle \Psi_{\mathrm{GS},+} |e^{+ i \theta \vec{n} \cdot \vec{S}_{\mathrm{tot}}} \vec{S}_{\mathrm{imp}} e^{- i \theta \vec{n} \cdot \vec{S}_{\mathrm{tot}}} | \Psi_{\mathrm{GS},+} \rangle = \langle \Psi_{\mathrm{GS},+} |e^{+ i \theta \vec{n} \cdot \vec{S}_{\mathrm{imp}}} \vec{S}_{\mathrm{imp}} e^{- i \theta \vec{n} \cdot \vec{S}_{\mathrm{imp}}} | \Psi_{\mathrm{GS},+} \rangle.
\end{align}
The norm of this expectation value is expressed as
\begin{align}
	|\langle \psi | \vec{S}_{\mathrm{imp}} | \psi \rangle| = \sqrt{\sum_{\alpha = x,y,z} \left| \langle \Psi_{\mathrm{GS},+} |e^{+ i \theta \vec{n} \cdot \vec{S}_{\mathrm{imp}}} S_{\mathrm{imp}}^{\alpha} e^{- i \theta \vec{n} \cdot \vec{S}_{\mathrm{imp}}} | \Psi_{\mathrm{GS},+} \rangle \right|^{2}}. \label{General_Snorm}
\end{align}
We consider a rotation matrix $R$ whose elements are $R^{\alpha \beta} = 2 \mathrm{Tr}\left[S_{\mathrm{imp}}^{\alpha} e^{i \theta \vec{n} \cdot \vec{S}_{\mathrm{imp}}} S_{\mathrm{imp}}^{\beta} e^{- i \theta \vec{n} \cdot \vec{S}_{\mathrm{imp}}}\right]$.
Utilizing the transformation properties of Pauli matrices under rotations, it follows that the matrix $R$ belongs to the special orthogonal group $\mathrm{SO}(3)$. 
Consequently, the action of the rotation matrix on the spin operators is described as 
\begin{align}
	e^{i \theta \vec{n} \cdot \vec{S}_{\mathrm{imp}}} S_{\mathrm{imp}}^{x} e^{- i \theta \vec{n} \cdot \vec{S}_{\mathrm{imp}}} &= R^{xx} S_{\mathrm{imp}}^{x} + R^{yx} S_{\mathrm{imp}}^{y} + R^{zx} S_{\mathrm{imp}}^{z}, \\ 
	e^{i \theta \vec{n} \cdot \vec{S}_{\mathrm{imp}}} S_{\mathrm{imp}}^{y} e^{- i \theta \vec{n} \cdot \vec{S}_{\mathrm{imp}}} &= R^{xy} S_{\mathrm{imp}}^{x} + R^{yy} S_{\mathrm{imp}}^{y} + R^{zy} S_{\mathrm{imp}}^{z}, \\
	e^{i \theta \vec{n} \cdot \vec{S}_{\mathrm{imp}}} S_{\mathrm{imp}}^{z} e^{- i \theta \vec{n} \cdot \vec{S}_{\mathrm{imp}}} &= R^{xz} S_{\mathrm{imp}}^{x} + R^{yz} S_{\mathrm{imp}}^{y} + R^{zz} S_{\mathrm{imp}}^{z}.
\end{align}
Given the orientation of $|\Psi_{\mathrm{GS},+} \rangle$, and noting that the expectation values $\langle \Psi_{\mathrm{GS},+} | S_{\mathrm{imp}}^{x} | \Psi_{\mathrm{GS},+} \rangle = \langle \Psi_{\mathrm{GS},+} | S_{\mathrm{imp}}^{y} | \Psi_{\mathrm{GS},+} \rangle = 0$ and $|\langle \Psi_{\mathrm{GS},+} | S_{\mathrm{imp}}^{z} | \Psi_{\mathrm{GS},+} \rangle| = \mathcal{S}$, we find from Eq.~\eqref{General_Snorm} that
\begin{align}
	\label{Sinvariance}
	|\langle \psi | \vec{S}_{\mathrm{imp}} | \psi \rangle| = \sqrt{|R^{zx} \mathcal{S}|^{2} + |R^{zy} \mathcal{S}|^{2} + |R^{zz} \mathcal{S}|^{2}} = \mathcal{S},
\end{align}
where we use $(R^{zx})^{2} + (R^{zy})^{2} + (R^{zz})^{2} = 1$ because $R \in \mathrm{SO}(3)$.
This gives Eq.~(3) of the main text.

\subsection{Derivation of Eq.~(4): Ground-state entanglement in the local moment phase}
We derive Eq.~(4) of the main text.
For the purpose, we find the form of $\mathcal{S}$ in the limit of small Kondo coupling, using the renormalization group (RG) analysis~\cite{Moca_supp}.
The Hamiltonian Eq.~\eqref{KondoHam} is written in energy-basis representation as 
\begin{align}
H = \sum_{\sigma = \uparrow,\downarrow} \int_{-D_{0}}^{D_{0}} d\epsilon \, \epsilon c_{\epsilon \sigma}^{\dagger} c_{\epsilon \sigma} + J \sum_{\sigma,\sigma' = \uparrow,\downarrow} \int_{-D_{0}}^{D_{0}} d\epsilon \, \int_{-D_{0}}^{D_{0}} d\epsilon' \, \sqrt{\rho(\epsilon) \rho(\epsilon')} c_{\epsilon \sigma}^{\dagger} \frac{\vec{\sigma}_{\sigma \sigma'}}{2} c_{\epsilon' \sigma'} \cdot \vec{S}_{\mathrm{imp}} + h S_{\mathrm{imp}}^{z}.
\end{align}
We include the last term of a local magnetic field $h$ at the impurity site and drop the potential scattering term.
$\rho(\epsilon) = \frac{1 + r}{2D_{0}^{1 + r}} |\epsilon|^{r}$ is the pseudogap DOS.
$c_{\epsilon \sigma}$ annihilates a bath electron with energy $\epsilon$ and spin $\sigma$.
We apply the decimation lowering the bare bandwidth $D_{0}$ to a running bandwidth $D = D_{0} e^{-l}$ with a scaling variable $l$. Then $j_{0} \equiv \rho(D_{0}) J$ is transformed to $j \equiv \rho(D) J(D)$. For small $J$ ($ \ll D_{0}$) and $r$, we find the perturbative RG equations~\cite{Moca_supp,Withoff_supp}
\begin{align}
\partial_{l} j &= - rj + j^{2} + \text{higher order terms}, \label{RGE1} \\
\partial_{l} \ln h &= - \frac{1}{2} j^{2} + \text{higher order terms}. \label{RGE2}
\end{align}
The integral of Eq.~\eqref{RGE1} gives
\begin{align}\label{RG integral for j}
\int_{0}^{l} dl = \int_{j_{0}}^{j(l)} \frac{1}{-rj + j^{2}} \, dj \qquad \Rightarrow \qquad j(l) = \frac{r}{1 - e^{rl} (1 - \frac{r}{j_{0}})}
\end{align}
which satisfies $j(l = 0) = j_{0}$ at $l = 0$. 
This shows that for $j_{0} > r$, $j(l)$ diverges at certain $l$ which corresponds the strongly coupled phase.
For $j_{0} < r$, $j(l)$ does not diverge in the RG process but $j(l) \rightarrow 0$ at $l \to \infty$ that corresponds to the local moment fixed point. The critical Kondo coupling strength $J_{\text{c}}$ is therefore determined by $\rho(D_{0}) J_{\text{c}} = r$.
With $\rho(D_{0}) = \frac{1+r}{2 D_{0}}$, we find $J_\text{c} = \frac{2r}{1+r} D_{0} \approx 2r D_{0}$~\cite{Withoff_supp}.
Next, the integral of Eq.~\eqref{RGE2} is combined with Eq.~\eqref{RG integral for j},
\begin{align}\label{hresult1}
\ln h(l = \infty) - \ln h(l = 0) = - \frac{1}{2} \int_{0}^{\infty} j(l')^{2} dl' = - \frac{1}{2} \int_{0}^{\infty} \Bigg[\frac{r}{1 - e^{rl'}(1 - \frac{r}{j_{0}})}\Bigg]^{2} dl' = \frac{j_{0}}{2} - \frac{r}{2} \ln \Bigg(\frac{r}{r - j_{0}}\Bigg).
\end{align}
We exponentiate both sides of Eq.~\eqref{hresult1},
\begin{align}
\frac{h(l = \infty)}{h(l = 0)} = e^{j_{0} / 2}\Bigg(\frac{r}{r - j_{0}}\Bigg)^{-r/2} 
= e^{j_{0} / 2}\Bigg(\frac{1}{1 - j_{0} / r}\Bigg)^{-r/2}.
\end{align}
We series-expand the result, considering the local moment phase,
\begin{align}\label{BfieldRelation}
\frac{h(l = \infty)}{h(l = 0)} = 1 - \frac{j_{0}^{2}}{4r} - \frac{j_{0}^{3}}{6r^{2}} + O(j_{0}^{4}).
\end{align}

We now compute $\langle \Psi_{\mathrm{GS},+} | S_{\mathrm{imp}}^{z} | \Psi_{\mathrm{GS},+} \rangle$, using Eq.~\eqref{BfieldRelation}.
It is written with the impurity's free energy~\cite{Moca_supp},
\begin{align}
	\langle \Psi_{\mathrm{GS},+} | S_{\mathrm{imp}}^{z} | \Psi_{\mathrm{GS},+} \rangle_{l = 0} = \lim_{h \rightarrow 0^{+}} - \frac{1}{k_{B}T} \frac{\partial}{\partial h} F_{\mathrm{imp}}(j_{0},h,D_{0}).
\end{align}
Using the chain rule, the free energy for the bare coupling and bandwidth at $l = 0$ relates with that of finite $l$,
\begin{align}\label{ChainRule}
	\frac{\partial}{\partial h} F_{\mathrm{imp}}(j_{0},h,D_{0}) = \Bigg(\frac{\partial h(l)}{\partial h}\Bigg) \frac{\partial}{\partial h(l)} F_{\mathrm{imp}}(j(l),h(l),D(l)).
\end{align}
With $\langle \Psi_{\mathrm{GS},+} | S_{\mathrm{imp}}^{z} | \Psi_{\mathrm{GS},+} \rangle_{l} = \lim_{h \rightarrow 0^{+}} - \frac{1}{k_{B}T} \frac{\partial}{\partial h(l)} F_{\mathrm{imp}}(j(l),h(l),D(l))$, this leads to
\begin{align} \label{SpinExpRelation}
	\langle \Psi_{\mathrm{GS},+} | S_{\mathrm{imp}}^{z} | \Psi_{\mathrm{GS},+} \rangle_{l = 0} = \left(\frac{\partial h(l = \infty)}{\partial h(l = 0)}\right) \langle \Psi_{\mathrm{GS},+} | S_{\mathrm{imp}}^{z} |\Psi_{\mathrm{GS},+} \rangle_{l = \infty},
\end{align}
Here $|\Psi_{\mathrm{GS},+} \rangle_{l}$ is the renormalized ground state with the spin quantum number $j_{\mathrm{tot}}^{z} = 1/2$ at $l$. 
At $l = \infty$ where no screening happens, 
this state becomes a product state $|\Psi_{\mathrm{GS},+} \rangle_{l = \infty} = |{\uparrow} \rangle \otimes |\psi_{\mathrm{bath}} \rangle$, where $|\psi_{\mathrm{bath}} \rangle$ is a quantum state in the bath having a spin quantum number $j_{\mathrm{tot}}^{z} = 0$. 
Using Eq.~\eqref{BfieldRelation} and $\langle \Psi_{\mathrm{GS},+} | S_{\mathrm{imp}}^{z} |\Psi_{\mathrm{GS},+} \rangle_{l = \infty} = \frac{1}{2}$,
we obtain
\begin{align} \label{SpinExpJc}
\langle \Psi_{\mathrm{GS},+} | S_{\mathrm{imp}}^{z} | \Psi_{\mathrm{GS},+} \rangle_{l=0} = \frac{1}{2}  \left(\frac{\partial h(l = \infty)}{\partial h(l = 0)}\right) = \frac{1}{2} \left[1 - \frac{j_{0}^{2}}{4r} - \frac{j_{0}^{3}}{6r^{2}} + O(j_{0}^{4})\right] = \frac{1}{2} - \frac{r}{8} \left(\frac{J}{J_{\text{c}}}\right)^{2} + O(J^{3}),
\end{align}
where $j_0 = J (1+r)/(2 D_{0}) = r J / J_\text{c}$ is used in the last equality.
Using $\mathcal{S} = \langle \Psi_{\mathrm{GS},+} | S_{\mathrm{imp}}^{z} | \Psi_{\mathrm{GS},+} \rangle_{l=0}$
[see the discussion below Eq.~\eqref{coefficients}]
and Eq.~\eqref{LMNformula}, for small $J\ll J_{\text{c}}$, we obtain Eq.~(4) of the main text,
\begin{equation}
	\mathcal{N}(T=0) = \frac{1}{2}r \left(\frac{J}{J_{\text{c}}}\right)^{2} + O((J/J_{\text{c}})^{3}).
\end{equation}

\subsection{Ground-state entanglement of the Kondo phase}
\label{sec:I1}

The ground state in the Kondo phase is nondegenerate and has the quantum numbers of $j_{\mathrm{tot}} = 0$ and $j_{\mathrm{tot}}^{z} = 0$, as it shares the same property with that in the diverging Kondo coupling limit.
Therefore, the entanglement negativity in the Kondo phase follows the known formula [Eq.~(4) of Ref.~\cite{Kim21_supp}],
\begin{align} \label{negativity_Kondo}
	\mathcal{N} = \sqrt{1 - 4 \mathcal{S}^{2}},
\end{align}
where $\mathcal{S} = |\langle \Psi^\text{K}_{\text{GS}} | \vec{S}_{\mathrm{imp}} |\Psi^\text{K}_{\text{GS}} \rangle|$ with the unique ground state $|\Psi^\text{K}_{\text{GS}}\rangle$.
The derivation of the formula is found in Supplementary Material of Ref.~\cite{Kim21_supp}.
This formula of the Kondo phase differs from the corresponding formula of the local moment phase in Eq. (3), reflecting the difference between the two fixed points (e.g., the ground-state degeneracy).

\subsection{Spin cloud length}
We derive the spin cloud lengths, $\xi_\text{K}$ (Kondo cloud length) and $\xi_\text{LM}$ (LM cloud length) in the Kondo and local-moment (LM) phases, respectively, by the pseudogap DOS.
From the RG equations~\eqref{RGE1} and~\eqref{RG integral for j}, we have 
\begin{equation}
	\label{universalscale}
	\frac{1}{r}
	\left(
	\ln\left|1-\frac{r}{j(l)}\right|-\ln\left|1-\frac{r}{j_{0}}\right|
	\right)
	= l = \ln\frac{D_{0}}{D} \quad \Rightarrow \quad 
	D_{0}\left|1-\frac{r}{j_{0}}\right|^{1/r}=D\left|1-\frac{r}{j(l)}\right|^{1/r},
\end{equation}
which is valid for $r \ll 1$.
This implies that the energy scale $D_{0}|1- r / j_0|^{1/r}$ [$= D_{0}|1- J_{\text{c}} / J|^{1/r}$ as $j_0 = J (1+r)/(2 D_{0}) = r J / J_\text{c}$; see Eq.~\eqref{SpinExpJc}] is invariant under the RG flow~\cite{Hewson97_supp}. 
This provides the characteristic energy scale of $D_{0}|1- J_{\text{c}} / J|^{1/r}$ in both the Kondo and LM phases, since
Eq.~\eqref{universalscale} is valid in both the phases. And the inverse of the energy, $|1-J_{\text{c}}/J|^{-1/r}/D_{0}$ provides the characteristic length scale in both the phases.

This can also be seen in the following approach based on the scaling hypothesis.
The spatial distribution of the spin cloud can be characterized by a function $C(L D_{0}, J \rho(D_0))$, where $L D_{0}$ and $J \rho(D_0)$ are the dimensionless length (a distance from the impurity) and energy.
The scaling hypothesis suggests $C(LD_{0}, J \rho(D_0)) \propto C(L D_0 e^{-l}, j(l))$ in the decimation of $D = D_{0} e^{-l}$ and $j(l) \equiv \rho(D) J(D)$ [see Eqs.~\eqref{RGE1} and~\eqref{RG integral for j}].
There is a special value of $l = l^{*}$ with which $|1- r / j (l^{*})|^{1/r} = 1$. 
This happens such that $j (l^{*})$ diverges in the Kondo phase ($j(l) > r$) and that $j (l^{*}) = r/2$ in the local moment phase ($j(l) < r$). At $l = l^{*}$, $C$ is determined by $L D_0 e^{-l^{*}}$.  This implies that $e^{l^*}/ D_0$, with $l^*$ satisfying $|1- r / j (l^{*})|^{1/r} = 1$, behaves as the characteristic length in both the phases.
Therefore, the Kondo cloud length is found as $\xi_{\text{K}} = e^{l^{*}}/D_{0} = |1-J_{\text{c}}/J|^{-1/r}/D_{0}\propto |1-J_c / J|^{-1/r}$ in the Kondo phase, and the LM cloud length also follows the same form $\xi_{\text{LM}} = e^{l^{*}}/D_{0} \propto |1- J_{\text{c}}/J|^{-1/r}$.

While Eq.~\eqref{universalscale} is valid for $r\ll1$, our numerical results show that the cloud has the length scale proportional to $|1-J_{\text{c}}/J|^{-\nu}$ with exponent $\nu$ for all studied values of $r$ in both the phases (see Figs.~\ref{Sfig_cloudscaling_SC_N} and \ref{Sfig_cloudscaling_LM_N}). The exponent $\nu$ only depends on $r$ ($\nu = 1/r$ at $r \ll 1$) and has the same value in both the phases for a given $r$. 
This is consistent with the fact that the correlation length $\xi$ diverges as $\xi\propto |1-J_{\text{c}}/J|^{-\nu}$  near the quantum critical point of $J=J_{\text{c}}$~\cite{Vojta06_supp}.
 
\section{II. Fixed point Analysis: Derivation of Eqs. (5) and (6)}
\label{sec:II}

To derive Eqs.~(5) and~(6) of the main text and find the $\alpha$ value in TABLE I of the main text for 
the fixed points (Kondo and local moment, LM) of the pseudogap DOS $\rho(\epsilon) = \frac{1 + r}{2D_0^{1 + r}} |\epsilon|^{r}$ and for the asymmetric local moment (ALM) fixed point of the diverging DOS, we develop an approach based on the fixed-point analysis of NRG.

We start with the Hamiltonian for the Wilson chain of finite length $N$~\cite{Wilson75_supp},
\begin{equation}\label{Wilsonchain}
	H_{N} = J\vec{S}_\text{imp} \cdot \sum_{\sigma,\sigma' = \uparrow,\downarrow}f_{0\sigma}^\dagger \frac{\vec{\sigma}_{\sigma\sigma'}}{2}f_{0\sigma'}+V\sum_{\sigma = \uparrow,\downarrow}f_{0\sigma}^\dagger f_{0\sigma}
	+\sum_{\sigma = \uparrow,\downarrow}\sum_{n=0}^{N-1} t_n\left(f_{n+1\sigma}^{\dagger}f_{n\sigma}+f_{n\sigma}^{\dagger}f_{n+1\sigma}\right),
\end{equation}
where $V$ is the potential scattering strength, $f_{n\sigma}$ is the annihilation operator of the $n$th site of the chain, and $N$ is the number of sites in the chain. In the limit $N \to \infty$, Eq.~\eqref{Wilsonchain} becomes identical to the original Hamiltonian in Eq.~\eqref{HWilsonchain}; as we consider the DOS satisfying $\rho(-\epsilon)=\rho(\epsilon)$, the on-site energy $\epsilon_{n}=0$ in Eq.~\eqref{HWilsonchain}. Wilson's logarithmic discretization makes the conduction electrons be described by a tight binding chain with the hopping strength $t_n\sim\Lambda^{-n/2}$, where $\Lambda$ is the discretization parameter.
For the convention~\cite{Wilson75_supp},~\cite{Buxton_supp}, we consider rescaling,
\begin{equation} \label{rescaledHN}
	H_{N} = \Lambda^{-(N-1)/2}\widetilde{H}_{N}.
\end{equation}
Since $t_n\sim\Lambda^{-n/2}$, the lowest hopping strength is order 1 in the rescaled form. 

We compute the matrix elements of the impurity spin operator $\vec{S}_{\mathrm{imp}}$. For the purpose, we calculate the eigenstates of Eq.~\eqref{Wilsonchain}, considering a fixed point Hamiltonian $H_{N}^{\mathrm{(FP)}}$. Since the Wilson chain Hamiltonian $H_{N}$ is close to $H_{N}^{\mathrm{(FP)}}$ for large $N$, Eq.~\eqref{Wilsonchain} can be written as the sum of the fixed point Hamiltonian and some perturbation terms $\{\delta H_{i}\}$~\cite{Wilson75_supp},
\begin{equation}\label{H perturbation form}
	H_{N}=H^{\text{(FP)}}_{N}+\sum_{i} \delta H_{i}.
\end{equation}
For a given fixed point, possible perturbation terms $\{\delta H_{i}\}$, corresponding to irrelevant operators, are found, based on symmetries of the Hamiltonian.
Then the eigenstates $| \psi \rangle$ of $H_N$ are written as
\begin{equation}\label{psi perturbation form}
	| \psi \rangle = | \psi^{\mathrm{(FP)}} \rangle + \sum_{i}|\delta \psi_{i} \rangle,
\end{equation}
where $| \psi^{\mathrm{(FP)}} \rangle$ is the corresponding eigenstate of the fixed point Hamiltonian and $|\delta \psi_{i} \rangle$ is the perturbation correction by $\delta H_{i}$.
We compute the matrix elements of the impurity spin operator $\vec{S}_{\mathrm{imp}}$,
\begin{equation}\label{Simp perturbation form}
	\langle \psi |\vec{S}_{\mathrm{imp}}| \psi \rangle
	= \langle \psi^{\mathrm{(FP)}} |\vec{S}_{\mathrm{imp}} | \psi^{\mathrm{(FP)}} \rangle
	+\sum_{i}\left[\langle \psi^{\mathrm{(FP)}} |\vec{S}_{\mathrm{imp}} | \delta\psi_{i} \rangle+\mathrm{H.c.} \right]
	+\sum_{i,j}\langle \delta\psi_{i} |\vec{S}_{\mathrm{imp}} | \delta\psi_{j} \rangle.
\end{equation}
Counting the scaling order of $\langle \psi^{\mathrm{(FP)}} |\vec{S}_{\mathrm{imp}} | \delta\psi_{i} \rangle$ and $\langle \delta\psi_{i} |\vec{S}_{\mathrm{imp}} | \delta\psi_{j} \rangle$ and finding the leading order term, we derive
\begin{equation}
	\label{matrixelements_general}
	|\langle E_i|\vec{S}_\mathrm{imp}|E_j\rangle|-\mathcal{S} \propto E^{\alpha_{\mathcal{S}}},
\end{equation}
where $|E_{i}\rangle, |E_{j}\rangle$ are eigenstates of $H_N$ with energies $E_{i}, E_{j}\sim E$. Note that $|\langle \psi^{\mathrm{(FP)}} |\vec{S}_{\mathrm{imp}} | \psi^{\mathrm{(FP)}} \rangle|=\mathcal{S}$.
The power $\alpha_{\mathcal{S}}$ depends on the fixed point.
By using Eq.~\eqref{matrixelements_general}, we derive Eqs.~(5) and (6) of the main text.

\subsection{Kondo fixed point}

We apply the above strategy to the Kondo fixed point of the pseudogap DOS.
At the Kondo fixed point, the system can be decomposed into two subsystems, the spin singlet formed by the impurity spin and the nearest site $f_{0 \sigma}$, and the rest part. 
Then the rescaled form $\widetilde{H}^{\text{(K)}}_{N}$ of the Kondo fixed point Hamiltonian $H^{\text{(K)}}_{N}$ [see Eq.~\eqref{rescaledHN}] is written as
\begin{equation}
	\label{Kondo rescaled FP H}
	\widetilde{H}^{\text{(K)}}_{N} =
	\Lambda^{(N-1)/2}J\vec{S}_{\text{imp}}\cdot\vec{s}(0)+
	\Lambda^{(N-1)/2}\sum_{\sigma = \uparrow,\downarrow} \sum_{n=1}^{N-1}t_n\left(f_{n+1\sigma}^{\dagger}f_{n\sigma}+f_{n\sigma}^{\dagger}f_{n+1\sigma}\right),
\end{equation}
where $\vec{s}(0)=\sum_{\sigma,\sigma' = \uparrow,\downarrow}f_{0\sigma}^\dagger \frac{\vec{\sigma}_{\sigma\sigma'}}{2}f_{0\sigma'}$. The potential scattering vanishes, $V=0$, at the fixed point by the particle-hole symmetry. For the convention, we diagonalize the second bath term in Eq.~\eqref{Kondo rescaled FP H},
\begin{equation}
	\label{SC bath diagonalization}
	\Lambda^{(N-1)/2}\sum_{\sigma = \uparrow,\downarrow}\sum_{n=1}^{N-1}t_n
	\left(f_{n+1\sigma}^{\dagger}f_{n\sigma}+f_{n\sigma}^{\dagger}f_{n+1\sigma}\right)
	=\sum_{\sigma = \uparrow,\downarrow}\sum_l\eta_l\left(g_{l\sigma}^\dagger g_{l\sigma}+h_{l\sigma}^\dagger h_{l\sigma}\right),
\end{equation}
where $g_{l\sigma}$ are particle operators and $h_{l\sigma}$ are hole operators.
The energy eigenvalues $\eta_{l}$ are sorted in ascending order, i.e. $\eta_{l}\leq\eta_{l'}$ for $l<l'$. Since $t_n\sim\Lambda^{-n/2}$, the lowest eigenvalue of the bath term in Eq.~\eqref{SC bath diagonalization} is of order 1.
Each low-lying eigenstate of the fixed point Hamiltonian has the following form
\begin{equation}\label{SC singlet}
	|\psi^{\mathrm{(K)}}\rangle=\frac{\ket{\uparrow}\ket{f_{0\downarrow}}-\ket{\downarrow}\ket{f_{0\uparrow}}}{\sqrt{2}}
	\otimes{\ket{\text{bath}(g_{l\sigma}, h_{l\sigma})}}.
\end{equation}
$\ket{\uparrow}$ and $\ket{\downarrow}$ denotes the eigenstate of $S_\mathrm{imp}^{z}$, $|f_{0\sigma} \rangle$ is a single fermion state with spin $\sigma$ at the nearest Wilson chain site. $|\text{bath}(g_{l\sigma},h_{l\sigma}) \rangle$ is an eigenstate of $\sum_{\sigma = \uparrow,\downarrow}\sum_l \eta_l (g_{l\sigma}^\dagger g_{l\sigma}+h_{l\sigma}^\dagger h_{l\sigma})$.
We can write $f_{n\sigma}$ operators by using $g_{l\sigma}$ and $h_{l\sigma}$,
\begin{equation}\label{f in SC}
	f_{1\sigma}=\Lambda^{-(N-1)(1-r)/4}\sum_l\left[c_{l} (g_{l\sigma} + h_{l\sigma}^\dagger)\right], \quad
	f_{2\sigma}=\Lambda^{-(N-1)(3-r)/4}\sum_l\left[c_{l}'(g_{l\sigma} - h_{l\sigma}^\dagger)\right],
\end{equation}
where $c_{l}$ and $c_{l'}$ are order 1 coefficients for small $l$~\cite{Buxton_supp}. 
The $N$-dependent coefficients $\Lambda^{-(N-1)(1-r)/4}$ and $\Lambda^{-(N-1)(3-r)/4}$ in Eq.~\eqref{f in SC} are useful for finding the scaling of the matrix elements.
The $N$-dependent coefficients are obtained by diagonalizing the bath term of the Wilson chain [Eq.~\eqref{SC bath diagonalization}], and those of $f_{n \sigma}$ are the same as the ones of $f_{m \sigma}$ for $m=n$ (mod 2) and $m,n \neq 0$~\cite{Wilson75_supp, Buxton_supp}.

At large but finite $N$, the Hamiltonian $H_{N} =H^\text{(K)}_N+\sum_i \delta H_i$ can be decomposed into the fixed point Hamiltonian $H^\text{(K)}_N$ and the perturbations $\sum_{i} \delta H_{i}$.
The perturbations $\delta H_i$ have the same symmetries as the original Hamiltonian, (i) the particle-hole symmetry, (ii) the U(1) symmetry corresponding to the total charge, and (iii) the SU(2) symmetry.
The possible perturbation terms, which preserve the symmetries and break the spin singlet state between the impurity spin and the $f_{0 \sigma}$ site (the ground state of the fixed point Hamiltonian $H^\text{(K)}_N$), are listed below:
\begin{equation}\label{SC perturbation terms}
\begin{aligned}
	&\delta H_{1} = \delta_{1}\sum_{\sigma}(f_{0\sigma}^\dagger f_{1\sigma} + f_{1\sigma}^{\dagger} f_{0\sigma}),\\
	&\delta H_{2} = \delta_{2} \vec{S}_\text{imp}\cdot
	\sum_{\sigma\sigma'}f_{1\sigma}^{\dagger} \frac{\vec{\sigma}_{\sigma\sigma'}}{2}f_{1\sigma'},\\
	&\delta H_{3} = \delta_{3}\sum_{\sigma}(f_{0\sigma}^\dagger f_{2\sigma} + f_{2\sigma}^{\dagger} f_{0\sigma}),\\
	&\delta H_{4} = \delta_{4} \vec{S}_\text{imp}\cdot
	\sum_{\sigma\sigma'}f_{2\sigma}^{\dagger} \frac{\vec{\sigma}_{\sigma\sigma'}}{2}f_{2\sigma'}.
\end{aligned}
\end{equation}
We consider the eigenstate $| \psi \rangle = | \psi^{\mathrm{(K)}} \rangle + \sum_{i}|\delta \psi_{i} \rangle$ of $H_{N}=H^\text{(K)}_{N}+\sum_{i}\delta H_{i}$ and compute  the matrix element $\langle \psi |\vec{S}_{\mathrm{imp}}| \psi \rangle$.
Here $| \psi^{\mathrm{(K)}} \rangle$ is the eigenstate of $H^\text{(K)}_{N}$, and $|\delta \psi_{i} \rangle$ is the correction by the perturbation $H_i$.

For $| \delta \psi_{1} \rangle$, we write the perturbed eigenstates explicitly~\cite{SLeethesis_supp} by using Eq.~\eqref{f in SC},
\begin{equation}
	\label{SC H1 correction}
	\ket{\delta\psi_1}\simeq\frac{\delta_1}{-\frac{3}{4}J}
	\sum_{\sigma=\uparrow,\downarrow}\ket{\sigma}
	\left(|f_{0\uparrow}f_{0\downarrow}\rangle\otimes f_{1\sigma}\ket{\text{bath}(g_{l\sigma}, h_{l\sigma})}
	+(-1)^\sigma f_{1\sigma}^\dagger\ket{\text{bath}(g_{l\sigma}, h_{l\sigma})}\right).
\end{equation}
Since there is only a single $f_{0\sigma}$ excitation term in $|\psi^{\mathrm{(K)}} \rangle$ [see Eq. \eqref{SC singlet}] while $\ket{\delta\psi_1}$ involves the double excitations or no excitation of $f_{0\sigma}$,
we find $\langle \psi^{\mathrm{(K)}} |\vec{S}_{\mathrm{imp}} | \delta\psi_{1} \rangle = 0$.
Instead,  $\langle \delta\psi_{1} |\vec{S}_{\mathrm{imp}} | \delta\psi_{1} \rangle$ is nonzero,
\begin{equation}\label{psi1 second order SC}
	\langle \delta\psi_{1} |\vec{S}_{\mathrm{imp}} | \delta\psi_{1} \rangle \propto 
	[\Lambda^{-(N-1)(1-r)/4}]^2 = \Lambda^{-(N-1)(1-r)/2}
\end{equation}
for low-energy excited states. This is found from the followings: (i) $f_{1\sigma}$ is of the order of $\Lambda^{-(N-1)(1-r)/4}$ [see Eq.~\eqref{f in SC}], (ii) the low-energy excited states are constructed by $\ket{\text{bath}(g_{l\sigma}, h_{l\sigma})}$ with small $l$, and (iii) $c_{l}$ is order 1 for small $l$.

For $| \delta \psi_{2} \rangle$, we write the perturbed eigenstates explicitly by using Eq.~\eqref{f in SC},
\begin{equation}\label{SC H2 correction}
\begin{aligned}
	|&\delta\psi_2\rangle
	\simeq\frac{\delta_2}{-\frac{3}{4}J}
	\left[
	\frac{-\ket{\uparrow}\ket{f_{0\uparrow}}}{2\sqrt{2}}\otimes f_{1\downarrow}^\dagger f_{1\uparrow}\ket{\text{bath}(g_{l\sigma}, h_{l\sigma})}+\frac{\ket{\downarrow}\ket{f_{0\downarrow}}}{2\sqrt{2}}\otimes f_{1\uparrow}^\dagger f_{1\downarrow}\ket{\text{bath}(g_{l\sigma}, h_{l\sigma})}\right]\\
	&+\frac{\delta_2}{-\frac{3}{4}J}\frac{\ket{\uparrow}\ket{f_{0\downarrow}}+\ket{\downarrow}\ket{f_{0\uparrow}}}{2\sqrt{2}}
	\otimes
	\Lambda^{-(N-1)(1-r)/2}
	\sum_{\bar{\sigma}}\sum_{\bar{l}_g, \bar{l}_h}\sum_{\bar{l}_g', \bar{l}_h'}
	\frac{\text{sgn}(\bar{\sigma})}{2}(c_{\bar{l}_g}g_{\bar{l}_g\bar{\sigma}}^\dagger+c_{\bar{l}_h}h_{\bar{l}_h\bar{\sigma}})
	(c_{\bar{l}_g'}g_{\bar{l}_g'\bar{\sigma}}+c_{\bar{l}_h'}h_{\bar{l}_h'\bar{\sigma}}^\dagger)
	{\ket{\text{bath}(g_{l\sigma}, h_{l\sigma})}},
\end{aligned}
\end{equation}
where $\bar{l}_g$ and $\bar{l}_h'$ indices are for unoccupied $g$ and $h$ states in $|\text{bath}(g_{l\sigma}, h_{l\sigma})\rangle$, and $\bar{l}_g'$ and $\bar{l}_h$ indices are for occupied states. 
We find $\langle \psi^{\mathrm{(K)}} |\vec{S}_{\mathrm{imp}} | \delta\psi_{2} \rangle = 0$, since
there is no overlap of bath states between the third term in Eq. \eqref{SC H2 correction} and $|\psi^{\mathrm{(K)}} \rangle$ in Eq. \eqref{SC singlet}~\cite{Moca_supp} while the first and second terms in Eq. \eqref{SC H2 correction} have different impurity spin and $f_{0 \sigma}$ states from $|\psi^{\mathrm{(K)}} \rangle$. 
Instead, as in the case of Eq.~\eqref{psi1 second order SC}, we find
\begin{equation}
	\langle \delta\psi_{2} |\vec{S}_{\mathrm{imp}} | \delta\psi_{2} \rangle \propto
	[(\Lambda^{-(N-1)(1-r)/4})^2]^2 = \Lambda^{-(N-1)(1-r)}.
\end{equation}
This contribution is negligible in comparison with that in Eq.~\eqref{psi1 second order SC}.
We also find that the contributions of $| \delta\psi_{3} \rangle$ and $| \delta\psi_{4} \rangle$ to the matrix element
$\langle \psi |\vec{S}_{\mathrm{imp}}| \psi \rangle$ in comparison with that in Eq.~\eqref{psi1 second order SC},
since the $N$-dependent coefficient of $f_{2\sigma}$ is smaller than that of $f_{1\sigma}$ [Eq.~\eqref{f in SC}] so that $f_{2\sigma}$ gives a sub-dominant scaling order than $f_{1\sigma}$ for low-energy states. 

Therefore, near the Kondo fixed point, the matrix elements of $\bra{\psi}\vec{S}_\text{imp}\ket{\psi}$ follow
\begin{equation}
		\bra{\psi}\vec{S}_\text{imp}\ket{\psi}
		\simeq
		\langle\psi^{\mathrm{(K)}}|\vec{S}_\text{imp}|\psi^{\mathrm{(K)}}\rangle
		+\bra{\delta\psi_1}\vec{S}_\text{imp}\ket{\delta\psi_1}  \simeq 0+\delta_{1}'\Lambda^{-(N-1)(1-r)/2} \propto E^{1-r}
\end{equation}
with a small coefficient $\delta_{1}'$.
In the last equality, $\Lambda^{- (N-1)/2}$ is replaced by the energy scale $E$ of the lowest excited states of the bath [the $N-1$ sites of the Wilson chain; see Eq.~\eqref{SC bath diagonalization}]~\cite{Wilson75_supp}.
This implies that for two states $|E_i\rangle$ and $|E_j\rangle$ having the same order of the energies $E_i, E_j\sim E\sim \Lambda^{- (N-1)/2}$, the matrix elements follow the scaling of
\begin{equation}
	\label{Kondo impurity spin matrix elements}
	|\langle E_i|\vec{S}_\mathrm{imp}|E_j\rangle|\propto E^{1-r}.
\end{equation}
This result is confirmed in our NRG calculation in Fig.~\eqref{Sfig_Szopscaling}.
Note that $\mathcal{S}=0$ in the Kondo phase [cf. Eq.~\eqref{matrixelements_general}].

\subsection{LM fixed point}

In the LM fixed point, where the impurity and the bath are independent, the rescaled fixed point Hamiltonian is
\begin{equation}\label{LM rescaled FP H}
	\widetilde{H}_{N}^{\text{(LM)}} =
	\Lambda^{(N-1)/2}\sum_{\sigma = \uparrow,\downarrow}\sum_{n=0}^{N-1}t_n\left(f_{n+1\sigma}^{\dagger}f_{n\sigma}+f_{n\sigma}^{\dagger}f_{n+1\sigma}\right).
\end{equation}
We write the eigenstates and eigenenergies by diagonalizing the rescaled form in Eq.~\eqref{LM rescaled FP H}.
\begin{equation}
	\Lambda^{(N-1)/2} \sum_{\sigma = \uparrow,\downarrow} \sum_{n=0}^{N-1} t_n
	\left(f_{n+1 \sigma}^{\dagger}f_{n \sigma}+f_{n \sigma}^{\dagger}f_{n+1 \sigma}\right)
	= \sum_{\sigma = \uparrow,\downarrow} \sum_l \eta_l' \left(g_{l\sigma}^\dagger g_{l\sigma} + h_{l\sigma}^\dagger h_{l\sigma}\right),
\end{equation}
where the lowest eigenenergy $\eta_{1}'$ has order 1 since we diagonalize the rescaled Hamiltonian.
Each low-lying eigenstate of Eq.~\eqref{LM rescaled FP H} is equivalent with an eigenstate of $\sum_l \eta_l'(g_{l \sigma}^\dagger g_{l\sigma} + h_{l\sigma}^\dagger h_{l\sigma})$, which is denoted as
\begin{equation}\label{LM psi0}
	|\psi^{\mathrm{(LM)}}\rangle = \ket{\chi}\otimes|\text{bath}(g_{l\sigma}, h_{l\sigma}) \rangle,
\end{equation}
where $\ket{\chi}$ is an arbitrary impurity state.
Similar to Eq.~\eqref{f in SC}, $f_{n\sigma}$ in the $g_{l\sigma}$ and $h_{l\sigma}$ basis are
\begin{equation}\label{f in LM}
	f_{0\sigma}=\Lambda^{-(N-1)(1+r)/4}\sum_l\left[c_l(g_{l\sigma}+h_{l\sigma}^\dagger)\right], \quad
	f_{1\sigma}=\Lambda^{-(N-1)(3+r)/4}\sum_l\left[c_l'(g_{l\sigma}-h_{l\sigma}^\dagger)\right],
\end{equation}
where $c_{l}$ and $c_{l}'$ are also order 1 constant for small $l$~\cite{Buxton_supp}.
Note that the $N$-dependent coefficients $\Lambda^{-(N-1)(1+r)/4}$ and $\Lambda^{-(N-1)(3+r)/4}$ are different from the Kondo fixed point case in Eq.~\eqref{f in SC}. In the LM fixed point, $f_{n\sigma}$ has the same $N$-dependent coefficient as $f_{m\sigma}$ when $m=n$ (mod 2).

For the same analogy with Eq.~\eqref{H perturbation form}, we write
\begin{equation}\label{LMwithperturbation}
	H_N= H_N^{\text{(LM)}} + \sum_i \delta H_i.
\end{equation}
Unlike Eq.~\eqref{Kondo rescaled FP H}, only the bath term exists in Eq.~\eqref{LM rescaled FP H}.
Therefore, Eq.~\eqref{LMwithperturbation} reads
\begin{equation}
	\label{LMwithperturbation_rescaling}
	H_{N}=\Lambda^{-(N-1)/2}
	\left[
	\sum_{\sigma = \uparrow,\downarrow} \sum_l \eta_l'\left(g_{l \sigma}^\dagger g_{l\sigma} + h_{l\sigma}^\dagger h_{l\sigma}\right)
	+\Lambda^{(N-1)/2}\sum_i \delta H_i
	\right],
\end{equation}
and note that the perturbation is rescaled as $\Lambda^{(N-1)/2}\sum_{i} \delta H_{i}$, instead of $\sum_{i}\delta H_{i}$.
The possible perturbation terms satisfying the symmetries of the original Hamiltonian are :
\begin{equation}\label{LM perturbation terms}
\begin{aligned}
	&\delta H_{1} = \delta_{1} \vec{S}_\text{imp}\cdot
	\sum_{\sigma\sigma'}f_{0\sigma}^{\dagger} \frac{\vec{\sigma}_{\sigma\sigma'}}{2}f_{0\sigma'},\\
	&\delta H_{2} = \delta_{2}\sum_{\sigma}(f_{0\sigma}^\dagger f_{1\sigma} + f_{1\sigma}^{\dagger} f_{0\sigma}),\\
	&\delta H_{3} = \delta_{3} \vec{S}_\text{imp}\cdot
	\sum_{\sigma\sigma'}f_{1\sigma}^{\dagger} \frac{\vec{\sigma}_{\sigma\sigma'}}{2}f_{1\sigma'},\\
	&\delta H_{4} = \delta_{4}\sum_{\sigma}(f_{0\sigma}^\dagger f_{2\sigma} + f_{2\sigma}^{\dagger} f_{0\sigma}).	
\end{aligned}
\end{equation}
We consider the eigenstates $| \psi \rangle = | \psi^{\mathrm{(LM)}} \rangle + \sum_{i}|\delta \psi_{i}\rangle$ of Eq.~\eqref{LMwithperturbation}, where $| \psi^{\mathrm{(LM)}} \rangle$ is the eigenstate of $H_{N}^{(\text{LM})}$, and $|\delta \psi_{i}\rangle$ is the correction by the rescaled perturbation $\Lambda^{(N-1)/2}\delta H_{i}$.

For $|{\delta\psi_{1}}\rangle$, by using Eq. \eqref{f in LM} and $\delta H_{1}$ in Eq.~\eqref{LM perturbation terms}, we calculate the perturbed eigenstate of $\Lambda^{(N-1)/2}\delta H_{1}$,
\begin{equation}\label{LM H1 correction}
	|\delta\psi_1\rangle
	\simeq\frac{\delta_1}{2}\Lambda^{-(N-1)r/2}
	\sum_{\bar{\sigma}\bar{\sigma}'}\sum_{\bar{l}_g, \bar{l}_h}\sum_{\bar{l}_g', \bar{l}_h'}
	\frac{\vec{S}_\text{imp}}{\eta_l'-\eta_{\bar{l}}'}\cdot
	(c_{\bar{l}_g}g_{\bar{l}_g\bar{\sigma}}^\dagger+c_{\bar{l}_h}h_{\bar{l}_h\bar{\sigma}})
	\frac{\vec{\sigma}_{\sigma\sigma'}}{2}
	(c_{\bar{l}_g'}g_{\bar{l}_g'\bar{\sigma}'}+c_{\bar{l}_h'}h_{\bar{l}_h'\bar{\sigma}'}^\dagger)
	\ket{\chi}\otimes|\text{bath}(g_{l\sigma}, h_{l\sigma})\rangle,
\end{equation}
where $\bar{l}_g$ and $\bar{l}_h'$ indices are for unoccupied $g$ and $h$ states in $|\text{bath}(g_{l\sigma}, h_{l\sigma})\rangle$, $\bar{l}_g'$ and $\bar{l}_h$ indices are for occupied states, $\eta_l$ is the energy of the bath state, and $\eta_{\bar{l}}$ is the energy change by the indices $\bar{l}_g$, $\bar{l}_h$, $\bar{l}_g'$, and $\bar{l}_h'$.
Similar to Eq. \eqref{SC H2 correction}, since the overlap of bath states between Eq. \eqref{LM H1 correction} and Eq. \eqref{LM psi0} is zero, $\langle \psi^{\mathrm{(LM)}} |\vec{S}_{\mathrm{imp}} | \delta\psi_{1} \rangle=0$.
However, $\langle \delta\psi_{1} |\vec{S}_{\mathrm{imp}} | \delta\psi_{1}\rangle$ is non-zero, we find
\begin{equation}\label{psi1 second order LM}
	\langle \delta\psi_{1} |\vec{S}_{\mathrm{imp}} | \delta\psi_{1}\rangle \propto
	[\Lambda^{-(N-1)r/2}]^2 = \Lambda^{-(N-1)r}.
\end{equation}

In the same way, we can find $|\delta\psi_2\rangle$ by using Eq.~\eqref{f in LM}.
In this case, $\langle \psi^{\mathrm{(LM)}} |\vec{S}_{\mathrm{imp}} | \delta\psi_{2} \rangle$ is non-zero, 
\begin{equation}
	\label{psi2 first order LM}
	\langle \psi^{\mathrm{(LM)}} |\vec{S}_{\mathrm{imp}} | \delta\psi_{2} \rangle
	+\langle \delta\psi_{2} |\vec{S}_{\mathrm{imp}} | \psi^{\mathrm{(LM)}} \rangle
	\propto \Lambda^{-(N-1)(1+r)/2}.
\end{equation}
It is followed by a similar reason for Eq.~\eqref{psi1 second order SC}, but the orders of $f_{0\sigma}$ and $f_{1\sigma}$ in Eq.~\eqref{f in LM} in the LM fixed point differ from the Kondo fixed point.
Additionally, comparing the orders of $f_{0\sigma}$ and $f_{1\sigma}$ in Eq.~\eqref{f in LM} reveals that $|\delta\psi_{3} \rangle$ and $| \delta\psi_{4} \rangle$ contribute negligibly to the matrix element compared to $|\delta\psi_{1} \rangle$ and $|\delta\psi_{2}\rangle$.

Thus, near the LM fixed point, from Eqs.~\eqref{Simp perturbation form},~\eqref{psi1 second order LM}, \eqref{psi2 first order LM}, the matrix elements of $\bra{\psi}\vec{S}_\text{imp}\ket{\psi}$ follow
\begin{equation}
	\label{LM Simp matrix calc.}
	\begin{aligned}
			\bra{\psi}\vec{S}_\text{imp}\ket{\psi}
			&\simeq
			\langle\psi^{\mathrm{(LM)}}|\vec{S}_\text{imp}|\psi^{\mathrm{(LM)}}\rangle
			+\bra{\delta\psi_{1}}\vec{S}_\text{imp}\ket{\delta\psi_{1}}
			+\langle \psi^{\mathrm{(LM)}}|\vec{S}_\text{imp}\ket{\delta\psi_{2}}
			+\langle \delta\psi_{2}|\vec{S}_\text{imp}|\psi^{\mathrm{(LM)}}\rangle\\
			&\simeq\mathcal{S}+\delta_{1}'\Lambda^{-(N-1)r}+\delta_{2}'\Lambda^{-(N-1)(1+r)/2}\\
			&\simeq\mathcal{S}+\delta_{1}'E^{2r}+\delta_{2}'E^{1+r},\\
		\end{aligned}
\end{equation}
where $\delta_{1}'$ and $\delta_{2}'$ are small coefficients. The non-zero $\mathcal{S}$ value comes from the impurity degree of freedom $\ket{\chi}$.
From Eq. \eqref{LM Simp matrix calc.}, we find the scaling of the impurity spin matrix elements,
\begin{equation}
	\label{LM impurity spin matrix elements}
	|\langle E_i|\vec{S}_\mathrm{imp}|E_j\rangle|-\mathcal{S}\propto
	\begin{cases}
		E^{2r}& r<1, \\
		E^{1+r}& r\geq 1.
	\end{cases}
\end{equation}
This result is also confirmed in our NRG calculation in Fig.~\eqref{Sfig_Szopscaling}.

\subsection{ALM fixed point}
At the ALM fixed point, the Kondo coupling vanishes like the LM fixed point, but the potential scattering term diverges due to the diverging DOS~\cite{Mitchell13_supp}.
Therefore, the rescaled ALM fixed point Hamiltonian can be written by
\begin{equation}
	\label{ALM rescaled FP H}
	\widetilde{H}_N^{\text{(ALM)}}
	= \Lambda^{(N-1)/2}\sum_{\sigma = \uparrow,\downarrow}\sum_{n=1}^{N-1}
	t_n\left(f_{n+1\sigma}^{\dagger}f_{n\sigma}+f_{n\sigma}^{\dagger}f_{n+1\sigma}\right).
\end{equation}
Differently from Eq.~\eqref{LM rescaled FP H} for the LM fixed point, the $f_{0}$ site are decoupled from the bath, because of the diverging potential scattering.
Thus, the bath term is the same as the Kondo fixed point case in Eq.~\eqref{Kondo rescaled FP H}, and it is diagonalized in the same way,
\begin{equation}
	\Lambda^{(N-1)/2}\sum_{n=1}^{N-1}t_n
	\left(f_{n+1}^{\dagger}f_{n} + f_{n}^{\dagger}f_{n+1}\right)
	=\sum_{\sigma = \uparrow,\downarrow} \sum_l \eta_l \left(g_{l\sigma}^\dagger g_{l\sigma} + h_{l\sigma}^\dagger h_{l\sigma}\right),
\end{equation}
where $g_{l\sigma}$ are particle operators and $h_{l\sigma}$ are hole operators.
The energy eigenvalues $\eta_{l}$ are sorted in ascending order, i.e. $\eta_{l}\leq\eta_{l'}$ for $l<l'$. Each low-lying eigenstates of the ALM fixed point Hamiltonian is given by
\begin{equation}
	\label{ALM psi0}
	|\psi^{\mathrm{(ALM)}}\rangle
	= \ket{\chi}\otimes| 0 \rangle_{0}\otimes|\text{bath}(g_{l\sigma}, h_{l\sigma}) \rangle,
\end{equation}
where $| 0 \rangle_{0}$ is a vacuum state of $f_{0\sigma}$ site. We write $f_{n \sigma}$ operators as
\begin{equation}\label{f in ALM}
	f_{1\sigma}=\Lambda^{-(N-1)(1-r)/4}\sum_l\left[c_{l} (g_{l\sigma} + h_{l\sigma}^\dagger)\right], \quad
	f_{2\sigma}=\Lambda^{-(N-1)(3-r)/4}\sum_l\left[c_{l}'(g_{l\sigma} - h_{l\sigma}^\dagger)\right].
\end{equation}
$c_{l}$ and $c_{l}'$ are order 1 constant for small $l$~\cite{Buxton_supp}.
Equation~\eqref{f in ALM} has the same form as Eq.~\eqref{f in SC}, but with $r<0$ for the ALM fixed point.
$f_{n \sigma}$ has the same $N$-dependent coefficient as $f_{m \sigma}$ when $m=n$ (mod 2) except $n=0$. 

The only difference in the rescaled fixed point Hamiltonian between the LM (Eq.~\eqref{LM rescaled FP H}) and ALM (Eq.~\eqref{ALM rescaled FP H}) cases is whether the $f_{0}$ site is decoupled or not.
Thus, the possible perturbations $\{\delta H_{i}\}$ are similar to the LM fixed point,
\begin{equation}
	\label{ALM perturbation terms}
\begin{aligned}
	&\delta H_{1} = \delta_{1} \vec{S}_{\text{imp}}\cdot
	\sum_{\sigma\sigma'}f_{1\sigma}^{\dagger} \frac{\vec{\sigma}_{\sigma\sigma'}}{2}f_{1\sigma'}\\
	&\delta H_{2} = \delta_{2}\sum_{\sigma}(f_{1\sigma}^\dagger f_{2\sigma} + f_{2\sigma}^{\dagger} f_{1\sigma})\\
	&\delta H_{3} = \delta_{3} \vec{S}_{\text{imp}}\cdot
	\sum_{\sigma\sigma'}f_{2\sigma}^{\dagger} \frac{\vec{\sigma}_{\sigma\sigma'}}{3}f_{2\sigma'}\\
	&\delta H_{4} = \delta_{4}\sum_{\sigma}(f_{2\sigma}^\dagger f_{3\sigma} + f_{3\sigma}^{\dagger} f_{2\sigma})	
\end{aligned}
\end{equation}
As compared with the $N$-dependent coefficients for the LM (Eq.~\eqref{f in LM}) and the ALM (Eq.~\eqref{f in ALM}) fixed points, the only difference is the sign of $r$.
Therefore, the matrix elements of the impurity spin of the ALM fixed point are obtained by reversing the sign of $r$ in Eq.~\eqref{LM impurity spin matrix elements},
\begin{equation}
	\label{ALM impurity spin matrix elements}
	|\langle E_{i}|\vec{S}_\mathrm{imp}|E_{j}\rangle|-\mathcal{S}\propto E^{-2r}.
\end{equation}
We note that the ALM phase only appears for $-1<r<0$.

\subsection{Derivation of Eqs. (5) and (6)}

We derive the power-law spatial distribution of the spin cloud in Eq.~(5) in the main text by using the matrix elements $\langle E_{i}|\vec{S}_{\mathrm{imp}}|E_{j}\rangle$ in Eq.~\eqref{matrixelements_general}.
The spin cloud is identified by applying the LSB at distance $L$ from the impurity [Eq.~\eqref{LSBSC} and~\eqref{LSBLM}]. We consider the ground-state expectation value of the impurity spin operator,
$\mathcal{S} = | \langle \Psi_\text{GS} | \vec{S}_\text{imp} | \Psi_\text{GS} \rangle$ for a ground state $| \Psi_\text{GS} \rangle$ in the absence of the LSB
and $\mathcal{S}' (L) = | \langle \Psi_\text{GS}' (L) | \vec{S}_\text{imp} | \Psi_\text{GS}' (L) \rangle|$ for the perturbed ground state $| \Psi_\text{GS}' (L) \rangle|$ in the presence of the LSB, and
study the dependence of the difference $\mathcal{S}' (L) - \mathcal{S}$ on $L$.

The LSB applied at distance $L$ is equivalent with the LSB applied to the $n$th Wilson chain site of $\Lambda^{-n/2} \sim L$. Hence, the perturbed ground state $|\Psi'_{\mathrm{GS}}(L)\rangle$ in the presence of the LSB has an energy proportional to $1/L$. The overlap $\langle E|\Psi'_{\mathrm{GS}}(L)\rangle$ between the ground state and an excited state $|E\rangle$ with energy $E$ is maximal when $E\sim1/L$, and it vanishes when $E$ significantly differs from $1/L$.
This is accordance with the energy scale separation in the Wilson chain.
This implies $\mathcal{S}' (L) \sim |\langle E_{i}|\vec{S}_\mathrm{imp}|E_{j}\rangle|$, where $E_{i}, E_{j} \sim E \sim 1/L$. From the power law in Eq.~\eqref{matrixelements_general}, we find
\begin{equation}
	\label{generalscalingofS}
	\mathcal{S}'(L)-\mathcal{S} \sim |\langle E_{i}|\vec{S}_\mathrm{imp}|E_{j}\rangle|-\mathcal{S}\propto E^{\alpha_{\mathcal{S}}} \propto L^{-\alpha_{\mathcal{S}}}.
\end{equation}
The value of the exponent $\alpha_{\mathcal{S}}$ is found in Eqs.~\eqref{Kondo impurity spin matrix elements}, \eqref{LM impurity spin matrix elements},  \eqref{ALM impurity spin matrix elements} for each fixed point. Combining this with the entanglement negativity formula for each phase 
[the formula in Eq.~(3) for the local moment phases, and Eq.~\eqref{negativity_Kondo} for the Kondo phase],
we obtain the power law in Eq.~(5) of the main text (see also TABLE I in the main text).
 
The power-law thermal suppression of the negativity $\mathcal{N} (T)$ in Eq.~(6) of the main text is also
derived by the power law in Eq.~\eqref{matrixelements_general} and the entanglement negativity formula for each phase, following a previous work in Ref.~\cite{Kim21_supp}.

\section{III. Entanglement and spin cloud for pseudogap DOS: NRG Calculation}
\label{sec:II}
\subsection{NRG details}
For the pseudogap and diverging DOS, we compute $\mathcal{S}$ and the entanglement negativity $\mathcal{N}$ by using the NRG. For the NRG calculation, we map the model Hamiltonian Eq.~\eqref{KondoHam} to the semi-infinite Wilson chain~\cite{Wilson75_supp}
\begin{equation}
	\label{HWilsonchain}
	H_{\text{WC}}
	= J\vec{S}_\text{imp} \cdot \sum_{\sigma,\sigma' = \uparrow,\downarrow}
	f_{0\sigma}^\dagger \frac{\vec{\sigma}_{\sigma\sigma'}}{2}f_{0\sigma'}
	+V\sum_{\sigma = \uparrow,\downarrow}f_{0\sigma}^\dagger f_{0\sigma}
	+\sum_{\sigma = \uparrow,\downarrow}\sum_{n=0}^{\infty}
	\left[
	t_{n}\left(f_{n+1\sigma}^{\dagger}f_{n\sigma}+f_{n\sigma}^{\dagger}f_{n+1\sigma}\right)
	+\epsilon_{n}f_{n\sigma}^{\dagger}f_{n\sigma}
	\right]
\end{equation}
where $f_{n\sigma}$ is the annihilation operator of the $n$th site of Wilson chain and $V$ is the potential scattering strength.
The hopping strength $t_{n}$ and the on-site energy $\epsilon_{n}$ are determined by the DOS $\rho(\epsilon)$ of the conduction electron bath.
$t_{n}$ decreases exponentially with increasing $n$~\cite{Wilson75_supp}, and $\epsilon_{n}=0$ when the DOS satisfies $\rho(-\epsilon)=\rho(\epsilon)$.
The DOS has the form of $\rho(\epsilon)=\frac{1+r}{2D}|\epsilon/D|^{r}$~\cite{Withoff_supp} with $r>0$ (the pseudogap DOS) or $r<0$ (the diverging DOS).
The energy window of the bath is $\epsilon \in [-D,D]$, and we set $D=1$. 
To construct the Wilson chain, we use the discretization parameter $\Lambda = 4$, the number $N_{\text{keep}}$ 
of kept states as $500$, the number $n_{z}$ of $z$-averaging as $2$~\cite{Wilson75_supp}.
We use the QSpace tensor network library~\cite{Weichselbaum12_supp, Weichselbaum20_supp} to exploit the charge U(1) and spin U(1) symmetries. To compute the negativity in Figs.~1 and~2 in the main text, we apply the NRG method developed in Ref.~\cite{Shim18_supp}. To compute $\mathcal{S}$, we follow the NRG in Ref.~\cite{Kim21_supp}.

\subsection{Local symmetry breaking}
We quantify the spatial distribution of spin clouds by computing the change $\delta\mathcal{N}(L) = |\mathcal{N} - \mathcal{N}'(L)|$ of the entanglement negativity (between the impurity and the conduction electrons) caused by a local perturbation on the conduction electrons at distance $L$ from the impurity. Here $\mathcal{N}$ and $\mathcal{N}'(L)$ are the negativities in the absence and presence of the perturbation. The quantification utilizes the fact that the entanglement more changes when more electrons participate in the cloud at distance $L$~\cite{Shim23_supp}.
The perturbation is referred as local symmetry breaking (LSB), as it breaks a symmetry of the system locally at the distance $L$. We also compute the quantity 
\begin{equation} 	\label{LSBSprime}
	\mathcal{S}'(L)=|\langle \Psi_{\text{GS}}'(L)|\vec{S}_{\text{imp}}|\Psi_{\text{GS}}'(L)\rangle|
\end{equation}
by using the NRG, to see the change of the quantity $\mathcal{S}$ by the LSB. Here $|{\Psi_{\text{GS}}'(L)}\rangle$ is a ground state in the presence of the LSB.
$\mathcal{N}'(L)$ and $\mathcal{S}'(L)$ obey the entanglement formula in Eq.~(3) of the main text.

Our choice of an LSB depends on the phases. For the Kondo phase, we choose a local magnetic field as the LSB,
\begin{equation}
	\label{LSBSC}
	H_{\text{LSB}}^{\text{(K)}}(L)
	= \delta B\frac{f_{n\uparrow}^{\dagger}f_{n\uparrow}-f_{n\downarrow}^{\dagger}f_{n\downarrow}}{2},
\end{equation}
where $\delta B$ is the strength of the small magnetic field.
It is applied to the $n$th Wilson chain site with $L=\Lambda^{n/2}/D$ (with the natural unit and the Fermi wave vector as 1), since the $n$th site represents the conductance electron region at distance $L$ from the impurity~\cite{Wagner18_supp, Krishna80_supp}. This LSB breaks the SU(2) spin symmetry at $L$.

For the local moment (LM) phase, we choose a local potential scattering term as the LSB,
\begin{equation}
	\label{LSBLM}
	H_{\text{LSB}}^{\text{(LM)}}(L)
	= \delta W(f_{n\uparrow}^{\dagger}f_{n\uparrow}+f_{n\downarrow}^{\dagger}f_{n\downarrow}),
\end{equation}
where $\delta W$ is the strength of the small potential scattering. This LSB breaks the particle-hole symmetry.
This LSB is chosen for the LM phase [rather than the local magnetic field in Eq.~\eqref{LSBSC}], since it does not break the SU(2) symmetry and hence it preserves the two-fold ground-state degeneracy, a central property of the LM phase.
For the asymmetric local moment (ALM) phase, we also employ the LSB in Eq.~\eqref{LSBLM}.

\subsection{NRG supporting Eq. (5) and Fig.~\ref{Sfig_cloudscaling_LM_N}: Scaling of entanglement in LM phase near the critical point}
\begin{figure}[h]
	\centerline{\includegraphics[scale=0.75]{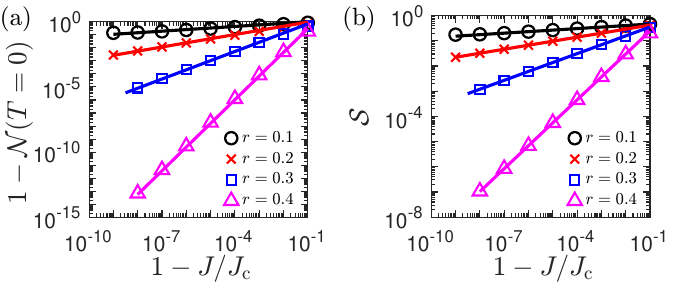}}
	\caption{
		NRG results (points) of (a) the entanglement negativity and (b) $\mathcal{S}$ in the LM phase near the critical point at zero temperature $T=0$. They agree with our prediction of the relation $1-\mathcal{N}\propto (1-J/J_\mathrm{c})^{2\beta}$ (lines) derived by combining Eq.~(3) of the main text and the known result $\mathcal{S}\propto (1-J/J_\mathrm{c})^{\beta}$ in (a) and the known expression~\cite{QimiaoSi_supp, Pixley15_supp} $\mathcal{S}\propto (1-J/J_\mathrm{c})^{\beta}$ (lines) in (b) for each $r$. 
		The exponent $\beta$ is 0.0603, 0.159, 0.353, 0.910 for $r=0.1$, 0.2, 0.3, 0.4, respectively. $\mathcal{N}$ and $\mathcal{S}$ are related by $\mathcal{N}(T=0)=\sqrt{1+4\mathcal{S}-8\mathcal{S}^2}-2\mathcal{S}$ ($\approx 1-6\mathcal{S}^2$ for $\mathcal{S}\ll 1/2$) in Eq.~(3) of the main text. 
	}
	\label{Sfig_exponentbeta}
\end{figure}

\subsection{NRG confirmation of Eq. (5): Spin cloud of the Kondo phase}
\begin{figure}[h]
	\centerline{\includegraphics[scale=0.7]{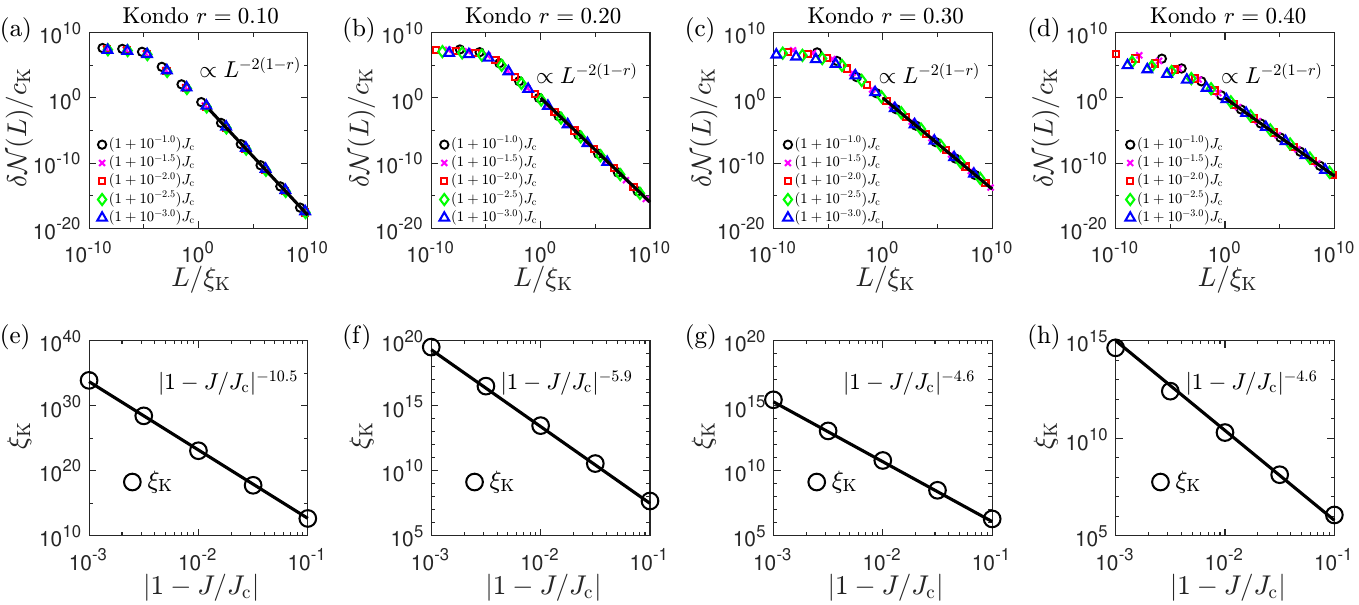}}
	\caption{NRG results of the spin cloud (upper pannels) and Kondo cloud length $\xi_{\text{K}}$ (lower pannel) of the Kondo phase of the psuedogap DOS with (a,e) $r=0.1$, (b,f) $0.2$, (c,g) $0.3$, (d,h) $0.4$. For different values of $J$ (indicated in the pannels) and $r$, the cloud follows the universal power law  $\delta\mathcal{N}(L)=c_{\text{K}} (L/\xi_{\text{K}})^{-2(1-r)}$ in Eq.~(5) in the main text. The exponent $2(1-r)$ agrees with the finding in TABLE I in the main text. The cloud length follows $\xi_{\text{K}}\propto|1- J / J_{\text{c}}|^{-\nu}$. In the panel (e) where $r \ll 1$, the exponent follows $\nu = 1/r$ as found in Eq.~\eqref{universalscale}. In the other pannels (f-h), it deviates from $\nu = 1/r$.
	}
	\label{Sfig_cloudscaling_SC_N}
\end{figure}

\newpage

\subsection{NRG confirmation of Eq. (5): Spin cloud of the local moment phase}
\begin{figure}[h]
	\centerline{\includegraphics[scale=0.69]{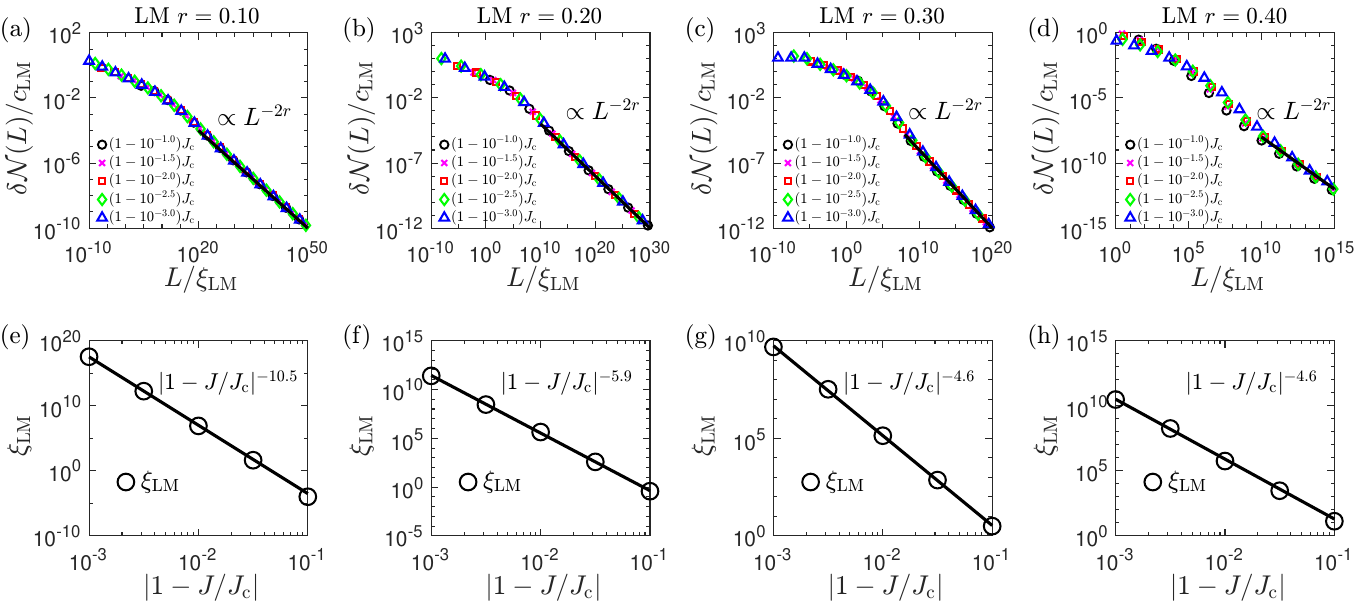}}
	\caption{NRG results of the spin cloud (upper pannels) and cloud length $\xi_{\text{LM}}$ (lower) of the local moment (LM) phase of the psuedogap DOS with (a,e) $r=0.1$, (b,f) $0.2$, (c,g) $0.3$, (d,h) $0.4$. 
		For different $J$'s (indicated in the pannels) and $r$'s, the cloud follows the universal power law  $\delta\mathcal{N}(L)=c_{\text{LM}}(L/\xi_{\text{LM}})^{-2r}$ with $c_{\text{LM}}\propto (1-J/J_\mathrm{c})^{\beta}$ in Eq.~(5) in the main text. The exponent $2r$ agrees with the finding in TABLE I in the main text. The cloud length follows $\xi_{\text{LM}}\propto|1- J/J_{\text{c}}|^{-\nu}$. 
		The exponent $\nu$ has the value of $\nu = 1/r$ at $r \ll 1$ [see Eq.~\eqref{universalscale}], and
		it has the same value as that of the Kondo phase for each $r$ [Fig.~\ref{Sfig_cloudscaling_SC_N}(e-h)]. $c_{\text{LM}} = \tilde{c}_{\text{LM}}(1-J/J_\mathrm{c})^{\beta}$ is found from (i) Eq.~(3) of the main text, which leads to $\delta \mathcal{N}(L) \approx 12 \mathcal{S}|\mathcal{S}-\mathcal{S}'(L)|$ near the critical point [$\mathcal{S}, \mathcal{S}'(L)\ll 1/2$; see Eq.~\eqref{LSBSprime}], (ii) the scaling behavior $\mathcal{S}\propto (1-J/J_\mathrm{c})^{\beta}$ in Fig.~\ref{Sfig_exponentbeta}(b), and (iii) the behavior $|\mathcal{S}-\mathcal{S}'(L)| \propto (L/\xi_{\text{LM}})^{-2r}$ in Eq.~\eqref{generalscalingofS}. $\tilde{c}_{\text{LM}}$ is independent of $J$ and $L$. 
	}
	\label{Sfig_cloudscaling_LM_N}
\end{figure}

\subsection{NRG confirmation of Eq. (6): Thermal suppression of entanglement in the Kondo phase}
\begin{figure}[h]
	\centerline{\includegraphics[scale=0.69]{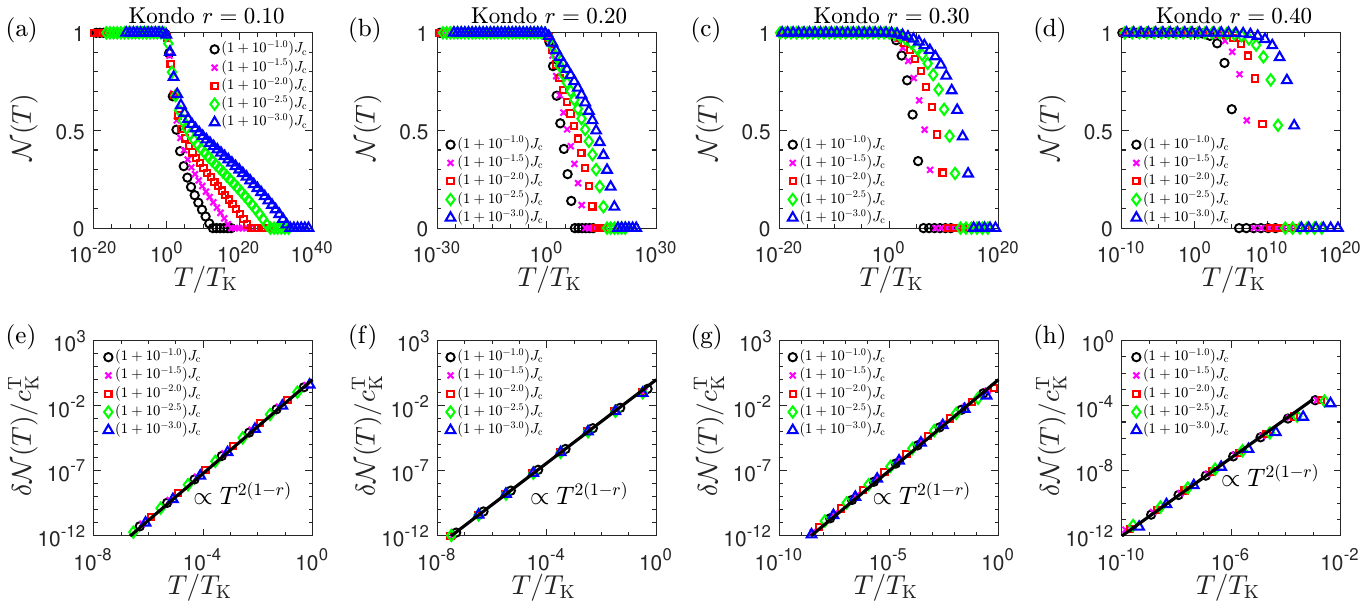}}
	\caption{NRG results of (a) the temperature dependence (upper panels) and
		(b) low-temperature suppression (lower) of the negativity $\mathcal{N}(T)$ for the Kondo phase of the psuedogap DOS with (a,e) $r=0.1$, (b,f) $0.2$, (c,g) $0.3$, (d,h) $0.4$. At temperature $T = 0$, $\mathcal{N} =  1$. The suppression follows the universal power law  $\delta\mathcal{N}(T)  = |\mathcal{N}(T=0)-\mathcal{N}(T)|=c_{\text{K}}^{\text{T}} (T/T_{\text{K}})^{2(1-r)}$ with constant $c_{\text{K}}^{\text{T}}$ in Eq.~(6) of the main text for different $J$'s and $r$'s. The Kondo temperature is the inverse of the cloud length, $T_{\text{K}} = 1/\xi_{\text{K}}$.}
	\label{Sfig_NTscaling_SC}
\end{figure}

\subsection{NRG confirmation of Eq. (6): Thermal suppression of entanglement in the local moment phase}
\begin{figure}[h]
	\centerline{\includegraphics[scale=0.70]{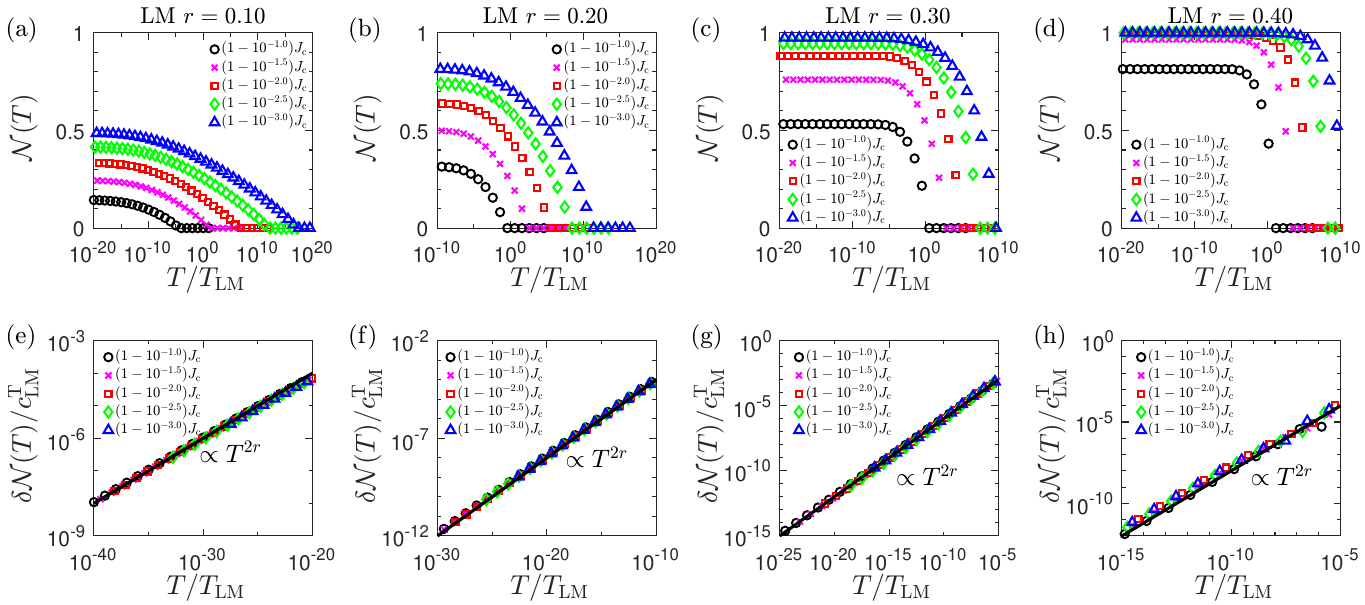}}
	\caption{NRG results of (a) the temperature dependence (upper panels) and
		(b) low-temperature suppression (lower) of the negativity $\mathcal{N}(T)$ for the local moment (LM) phase of the psuedogap DOS with (a,e) $r=0.1$, (b,f) $0.2$, (c,g) $0.3$, (d,h) $0.4$. At zero temperature $T=0$, $\mathcal{N}$ has a nonzero value smaller than 1. The suppression obeys the universal power law  $\delta\mathcal{N}(T)  = |\mathcal{N}(T=0)-\mathcal{N}(T)|=c_{\text{LM}}^{\text{T}} (T/T_{\text{LM}})^{2r}$ with $c_{\text{LM}}^{\text{T}}\propto (1-J/J_{\text{c}})^{\beta}$ in Eq.~(6) of the main text for different $J$'s and $r$'s. The LM temperature is the inverse of the LM cloud length, $T_{\text{LM}} = 1/\xi_{\text{LM}}$. The exponent $\beta$ of $c_{\text{LM}}^{\text{T}}$ is discussed in Fig.~\ref{Sfig_cloudscaling_LM_N}. }
	\label{Sfig_NTscaling_LM}
\end{figure}

\subsection{NRG confirmation of the analysis in Sec.~II: NRG data of the impurity spin matrix elements}
\begin{figure}[h]
	\centerline{\includegraphics[scale=0.7]{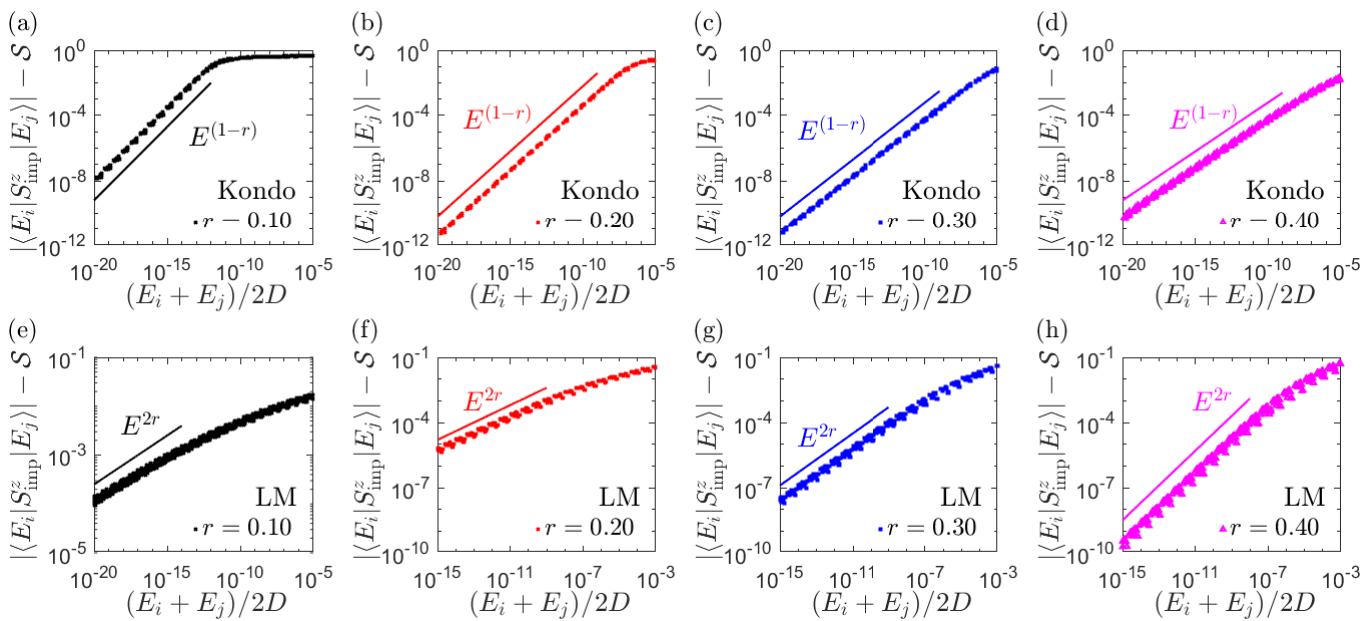}}
	\caption{NRG results of the matrix elements $|\langle E_{i}|\vec{S}_\mathrm{imp}|E_{j}\rangle|-\mathcal{S}\propto E^{\alpha_{\mathcal{S}}}$ for the Kondo phase (upper panels) and the local moment (LM) phase (lower) of the psuedogap DOS with (a,e) $r=0.1$, (b,f) $0.2$, (c,g) $0.3$, (d,h) $0.4$. The exponent $\alpha_{\mathcal{S}}$ is $1-r$ in the Kondo phase, $2r$ in the LM phase with $r<1$,
		$1+r$ in the LM phase with $r>1$ (not shown), and $-2r$ in the ALM phase (not shown)
		in agreement with Eqs.~\eqref{matrixelements_general}, \eqref{Kondo impurity spin matrix elements}, \eqref{LM impurity spin matrix elements}, \eqref{ALM impurity spin matrix elements}. In each NRG iteration step, we obtain the eigenstates $|E_{i}\rangle$ whose eigenenergies correspond to the NRG step, and plot the 200 largest matrix elements $|\langle E_{i}|\vec{S}_\mathrm{imp}|E_{j}\rangle|$  with the average energy $(E_{i}+E_{j})/2D$~\cite{Kim21_supp}.
	}
	\label{Sfig_Szopscaling}
\end{figure}

\section{IV. Entanglement and spin cloud for hard gap DOS}
\label{sec:IV}

\subsection{SSH Kondo model and DMRG}

To describe conduction electrons having a hard energy gap, we employ the semi-infinite Su-Schrieffer-Heeger (SSH) model.
The Hamiltonian can be written as $H = H_{\text{SSH}} + H_{\text{K}}$,
\begin{equation}
		H_{\text{SSH}} = \sum_{\sigma=\uparrow,\downarrow}\sum_{n=1}^N (t_{A}
		\psi^{\dagger}_{A,n,\sigma}\psi_{B,n,\sigma}
		+t_{B} \psi^{\dagger}_{B,n,\sigma}\psi_{A,n+1,\sigma}+\mathrm{H.c.}), \quad \quad
		H_{\text{K}}= J\sum_{\sigma\sigma'}
		\psi_{A,1,\sigma}^{\dagger}\frac{\vec{\sigma}_{\sigma\sigma'}}{2}\psi_{A,1,\sigma'}\cdot
		\vec{S}_{\text{imp}}.
\end{equation}
$H_{\text{SSH}}$ describes conduction electrons in the SSH lattice having alternating hopping energies $t_{A} = t+\Delta/2$ and $t_{B} = t-\Delta/2$. $t$ is the average hopping energy. $\Delta$ leads to the energy gap.
The Fermi velocity $v_{\mathrm{F}}=2t$. The gap correlation length $\xi_{\Delta} = v_{\mathrm{F}}/\Delta$.
$H_{\text{K}}$ shows the coupling between the impurity spin and the spin of the first site of the lattice.

To draw Fig.~4 of the main text, we obtain the ground states of the Hamiltonian $H$ 
in the presence of the additional LSB (the local potential scattering) at the $A$ site of $n=(L+1)/2$
by using the the density matrix renormalization group (DMRG) method,  
and compute the entanglement negativity between the impurity and the rest of the system in the ground state.
For the DMRG calculation, we consider $N=200$ lattice sites and the bond dimension 1000.
To improve the DMRG calculation, we use the QSpace tensor network library~\cite{Weichselbaum12_supp, Weichselbaum20_supp} to exploit the charge U(1) and spin SU(2) symmetries.  
The same parameters are used in the computation of the results in Figs.~\ref{Sfig_SSHSzszcorr_numerics} and \ref{Sfig_SSHSzszcorr}

For the SSH model, we confirm the relation between the entanglement negativity and the quantity $\mathcal{S} = |\langle \Psi_\text{GS} | \vec{S}_\text{imp} | \Psi_\text{GS} \rangle|$ in Eq.~(3) of the main text by using the DMRG results of $\mathcal{S}$ and the negativity.

\subsection{DMRG results: Spin-spin correlation}

\begin{figure}[h]
	\centerline{\includegraphics[scale=0.7]{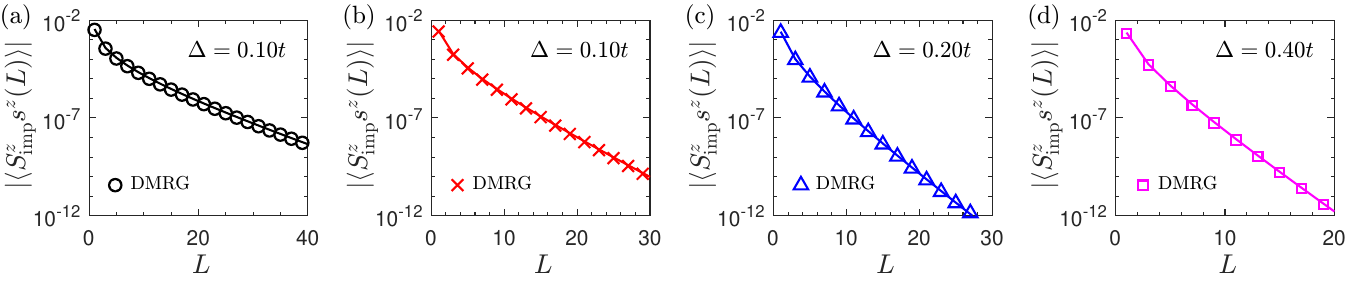}}
	\caption{DMRG results (points) of the correlation $\langle S_{\text{imp}}^{z} s^{z}(L)\rangle$ in the SSH model. The results agree with the perturbation theory (lines) in Eq.~\eqref{Sdots result}. The parameters are $J=0.1t$ and
		(a) $\Delta=0.10t$ (b) $\Delta=0.20t$ (c) $\Delta=0.30t$ (d) $\Delta=0.40t$.}
	\label{Sfig_SSHSzszcorr_numerics}
\end{figure}

In the main text, the exponential decay of the spin cloud as a function of the distance $L$ from the impurity is found by computing the entanglement negativity in the presence of the LSB at the distance.
The exponential decay also appears in the correlation $\langle S_{\text{imp}}^{z}s^{z}(L)\rangle$ between the impurity spin $S_{\text{imp}}^{z}$ and the spin $s^{z}(L)$ of conduction electrons at the distance $L$.
This is found in our DMRG results in Fig.~\ref{Sfig_SSHSzszcorr_numerics}, which
agree with our perturbation theory in Eq.~\eqref{Sdots result}.

In the main text, we show that for the hard gap DOS, the LM cloud exhibits the spatial exponential decay by a universal function $\propto e^{- 2L / \xi_\textrm{LM}}$ and that the LM cloud length $\xi_\textrm{LM}$ depends on competition between the bare Kondo cloud length $\xi_\text{K}^{(0)}$ (in the absence of the gap, $\Delta = 0$) and the gap correlation length $\xi_\Delta = v_\text{F} / \Delta$, $\xi_\textrm{LM} = \xi_\Delta$ when $\xi_\text{K}^{(0)} \ll \xi_\Delta$, and $\xi_\textrm{LM} = \xi_{\Delta}'$ when $\xi_\text{K}^{(0)} \ll \xi_\Delta$. $\xi_{\Delta}'$ has nontrivial dependence on $\Delta$ and the detailed DOS form.

\begin{figure}[h]
	\centerline{\includegraphics[scale=0.70]{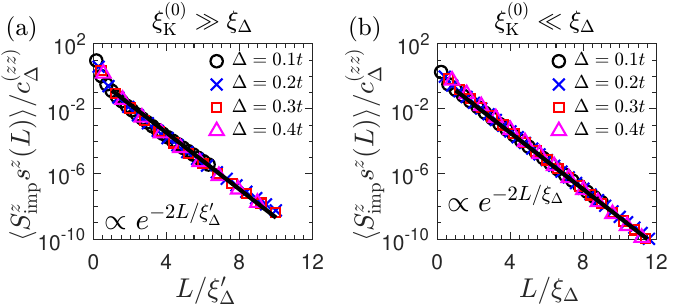}}
	\caption{DMRG results (points) of the correlation $\langle S_{\text{imp}}^{z} s^{z}(L)\rangle$ in
		the SSH model with various parameters. They lie on the single curve following the exponential decay (lines) in Eq.~\eqref{SdotsDMRG}, implying the universality. (a) The regime of  $\xi_{\text{K}}^{(0)}\gg\xi_{\Delta}$ with $J=0.1t$. $\xi_{\text{K}}^{(0)}$ is longer than $10^{5}$ lattice sites~\cite{Sorensen96_supp}. The LM cloud length is $\xi_{\Delta}'\approx 1.40, 1.35, 1.33, 1.32 v_{\mathrm{F}}/\Delta$ for $\Delta = 0.1, 0.2, 0.3, 0.4t$, respectively.
		(b) The regime of $\xi_{\text{K}}^{(0)}\ll\xi_{\Delta}$ with $J=16t$.  $\xi_{\text{K}}^{(0)}$ is shorter than $0.03$ lattice sites~\cite{Sorensen96_supp}. The LM cloud length is 
		$\xi_{\Delta}$.}
	\label{Sfig_SSHSzszcorr}
\end{figure}

The same universal exponential decay with the LM cloud length is also found in the correlation $\langle S_{\text{imp}}^{z}s^{z}(L)\rangle$,
\begin{equation}
	\label{SdotsDMRG}
	\langle S_{\text{imp}}^{z} s^{z}(L)\rangle = c_{\Delta}^{(zz)} e^{-2L/\xi_{\text{LM}}},
\end{equation}
according to our DMRG results in Fig.~\ref{Sfig_SSHSzszcorr}.
They support those in Fig.~4 of the main text.
In the regime of $\xi_{\text{K}}^{(0)}\gg \xi_{\Delta}$, they show that $\xi_\textrm{LM} = \xi_{\Delta}'$
and the constant $c_{\Delta}^{(zz)}\propto \Delta$.
When $\xi_{\text{K}}^{(0)}\ll\xi_{\Delta}$, they show that $\xi_\textrm{LM} = \xi_{\Delta}$ and $c_{\Delta}^{(zz)}\propto \Delta^{2}$.

\subsection{Perturbation theory: Spin-spin correlation}

We develop a perturbation theory where the Kondo coupling $J$ is used as the small perturbation parameter.
The theory is applicable to the regime of $\xi_{\text{K}}^{(0)}\gg \xi_{\Delta}$, and agrees with the DMRG results.
It shows that $\xi_\textrm{LM} = \xi_{\Delta}'$, providing a way to find $\xi_{\Delta}'$ numerically. 

To construct the bare states in the perturbation theory, we find the eigenstates of the semi-infinite SSH lattice,
\begin{equation}
\begin{aligned}
	\label{eigstates}
	c_{\pm,k, \sigma}
	&=\sum_{n=1}^{\infty}\phi_{\pm,A,n,k}\psi_{A,n,\sigma}
	+ \sum_{n=1}^{\infty}\phi_{\pm,B,n,k}\psi_{B,n,\sigma}, \\
	\phi_{\pm, A,n,k} &= \sqrt{\frac{1}{\pi}}\frac{1}{E_{k}}(t_{A}\sin nk+t_{B}\sin(n-1)k)
	=  \sqrt{\frac{1}{\pi}}\sin(nk-2\varphi),  \\
	\phi_{\pm, B,n,k} &= \pm \sqrt{\frac{1}{\pi}}\sin nk.
\end{aligned}
\end{equation}
The sign factor $\pm$ corresponds to the positive and negative energy eigenvalues. The energy eigenvalue $E_{k}$ and the phase factor $\varphi$ are determined by $E_{k}^{2} = t_{A}^{2}+t_{B}^{2}+2t_{A}t_{B}\cos k$ and $t_{A}+t_{B}e^{\pm ik} = E_{k}e^{\pm 2i\varphi}$. 
One of the bare ground state and energy are $|\Psi_{\text{GS},+}\rangle = \prod_{k,\sigma} c_{-,k,\sigma}^{\dagger}|0\rangle\otimes |\uparrow\rangle$ and $E_{\text{GS}} = -2\int_{0}^{\pi}dk E_{k}$. $|{0}\rangle$ is the vacuum state.

The first-order perturbation correction $O(J)$ of the ground state is found as
\begin{equation}
	|\delta\Psi_{\text{GS},+}^{(1)}\rangle
	= \frac{J}{4}\left(\frac{1}{\pi}\right)\int_{0}^{\pi}dk\int_{0}^{\pi}dk'
	\frac{\sin(k-2\varphi(k))\sin(k'-2\varphi(k'))}{E_{k}+E_{k'}}
	(c_{+,k,\uparrow}^{\dagger}c_{-,k',\uparrow}-c_{+,k,\downarrow}^{\dagger}c_{-,k',\downarrow})
	|\Psi_{\text{GS},+}\rangle.
\end{equation}
Using the correction, we derive the correlation between the impurity spin and the electron spin at the distance $L$ from the impurity at the $(L+1)/2$-th $A$ site,
\begin{equation}
\begin{aligned}
	\label{Sdots result}
	&\langle S_{\text{imp}}^{z} s^{z}(L)\rangle
	\approx \langle{\Psi_{\text{GS},+}}|S_{\text{imp}}^{z} s_{A, (L+1)/2}^{z}|{\Psi_{\text{GS},+}}\rangle
	+\langle{\delta\Psi_{\text{GS},+}^{(1)}}|S_{\text{imp}}^{z} s_{A, (L+1)/2}^{z}|{\Psi_{\text{GS},+}}\rangle
	+\langle{\Psi_{\text{GS},+}}|S_{\text{imp}}^{z} s_{A, (L+1)/2}^{z}|{\delta\Psi_{\text{GS},+}^{(1)}}\rangle\\
	&=-\frac{1}{t_{B}^{2}}\frac{J}{4}\left(\frac{1}{\pi}\right)^{2}
	\int_{t_{A}-t_{B}}^{t_{A}+t_{B}}dE\int_{t_{A}-t_{B}}^{t_{A}+t_{B}}dE'
	\frac{\sin(\frac{(L+1)}{2}\cos^{-1}(\frac{E^{2}-t_{A}^{2}-t_{B}^{2}}{2t_{A}t_{B}})-2\varphi(E))
	\sin(\frac{(L+1)}{2}\cos^{-1}(\frac{E'^{2}-t_{A}^{2}-t_{B}^{2}}{2t_{A}t_{B}})-2\varphi(E'))}{E+E'}.
\end{aligned}
\end{equation}
This expression agrees with our DMRG results in Figs.~\ref{Sfig_SSHSzszcorr_numerics} and \ref{Sfig_SSHSzszcorr}.
We compute the integral in Eq.~\eqref{Sdots result} numerically and find $\xi'_\Delta$ by fitting it to Eq.~\eqref{SdotsDMRG}.


\begin{thebibliography}{99}
\bibitem{defofS} $\mathcal{S}$ equals the quantity $\lim\limits_{h \rightarrow 0^{+}}\lim\limits_{T \rightarrow 0}\text{Tr}(\hat{\rho}(T, h)S_{\text{imp}}^{z})$ used in literature~\cite{Ingersent02,Pixley15}, where $h$ is a local magnetic field applied to the impurity spin.
\bibitem{LMdegeneracy} In LM phases, for an arbitrary ground state $|\Psi_{\mathrm{GS},+}\rangle$ having the expectation value $\langle\Psi_{\text{GS},+}|\vec{S}_{\mathrm{imp}}|\Psi_{\text{GS},+}\rangle$ of the impurity spin operator, there is the orthogonal degenerate ground state $|\Psi_{\mathrm{GS},-}\rangle$ satisfying $\langle\Psi_{\text{GS},-}|\vec{S}_{\mathrm{imp}}|\Psi_{\text{GS},-}\rangle = -\langle\Psi_{\text{GS},+}|\vec{S}_{\mathrm{imp}}|\Psi_{\text{GS},+}\rangle$ due to the SU(2) symmetry~\cite{Supp}.
\bibitem{LSB} For the LM and ALM phases, the LSB that we choose is a local potential scattering. This LSB breaks the particle-hole symmetry, but preserves the SU(2) symmetry that results in degenerate ground states, a feature of the LM and ALM phases. For the Kondo phase, the LSB is a local magnetic field, which breaks the SU(2) symmetry.
\bibitem{decaystozero} $\delta\mathcal{N}(L)$ vanishes as $L \rightarrow \infty$, contrary to the entanglement entropy that remains finite when the partition into subsystems for the entanglement is placed far away from the impurity (Ref.~\cite{Wagner18}).
\end{thebibliography}
\end{document}